\documentclass[aps,prc,showpacs,floatfix,groupedauthor,10pt,twocolumn,nofootinbib]{revtex4}

\newcommand{\ifproofpre}[2]{#1}

\usepackage{graphicx}
\usepackage{amsmath}
\usepackage{amssymb}
\usepackage{mathtools}
\usepackage{liealg}
\usepackage{wignert}
\usepackage{isotope}
\usepackage{dcolumn}
\usepackage{mathrsfs}

\DeclareRobustCommand{\abs}[1]{\lvert#1\rvert}

\DeclareRobustCommand{\grpsochain}{\grpso{5}\supset\grpso{3}}

\DeclareRobustCommand{\grpsutimes}{\grpsu{1,1}\times\grpso{5}}
\DeclareRobustCommand{\grpy}[2][]{\grp{Y}{#1}{#2}}

\DeclareRobustCommand{\deltasqrt}[1]{(1+\delta_{#1})^{1/2}}

\DeclareRobustCommand{\Deltahat}{\hat{\Delta}}
\DeclareRobustCommand{\gammabar}{\bar{\gamma}}

\DeclareRobustCommand{\Lambdahat}{\hat{\Lambda}}
\DeclareRobustCommand{\calM}{\mathcal{M}}
\DeclareRobustCommand{\calQ}{\mathcal{Q}}
\DeclareRobustCommand{\bbR}{\mathbb{R}}
\DeclareRobustCommand{\scrD}{\mathscr{D}}

\DeclareRobustCommand{\vmax}{{v_\text{max}}}

\newcommand{\fnPgamma}{\footnote{
The expression~(\ref{eqn-Pgamma-sum}) for
                  $P(\gamma)$ is equivalent to~(A7) of
                  Ref.~\cite{caprio2005:axialsep}.  However, the 
                  normalization factors appearing in these expressions differ,
                  due to the different
                  normalization conventions defined for the $F_K(\gamma)$
                  coefficients in
                  Ref.~\cite{caprio2005:axialsep} and in the present
                  work, which follows
                  Ref.~\cite{caprio2009:gammaharmonic}.
}}

\newcommand{\fngenpotl}{\footnote{
Any potential $V(\gamma)$ satisfying the basic requirements from the
Bohr coordinate symmetries may be expanded in terms of the form
$\cos^n3\gamma$ (which is equivalent to Fourier 
decomposition in terms of the form $\cos3n\gamma$) and therefore may readily be
accommodated for calculations within the ACM.  }}

\newcommand{\fnKnode}{\footnote{
When only one $K$ term contributes
to~(\ref{eqn-expansion-F}), as in~(\ref{eqn-intrinsic}), a zero-crossing in $F_K(\gamma)$
necessarily yields a zero-valued minimum in $P(\gamma)$.  If, instead, the
minimum is washed out, it may be concluded that multiple $K$ terms are contributing in~(\ref{eqn-Pgamma-F}),
such that these terms do not simultaneously have nodes at the same $\gamma$ value.
}}

\newcommand{\fntriaxortho}{\footnote{
Moreover, under adiabatic separation, the
                  $\gamma$ wave function for the excited $K=0$ band
                  must be orthogonal to the ground state band wave
                  function.  Since this distribution has moved to larger
                  $\gamma$ values, the redistribution in probability
                  to the smaller-$\gamma$ peak for the excited band can
                  be understood from orthogonality constraints, following
                  arguments similar to those applied in
                  Ref.~\cite{sato2010:triax-dynamics} for 
                  prolate-oblate coexistence.
}}

\begin{document}
%bibliographystyle{apsrevm}

%***************************************************************************
% front matter
%***************************************************************************

\title{\boldmath Exact diagonalization of the Bohr Hamiltonian for rotational nuclei:
Dynamical $\gamma$ softness and triaxiality}

\author{M. A. Caprio}
\affiliation{Department of Physics, University of Notre Dame,
Notre Dame, Indiana 46556-5670, USA}

\date{\today}

\begin{abstract}
  Detailed quantitative predictions are obtained for phonon and
  multiphonon excitations in well-deformed rotor nuclei within the
  geometric framework, by exact numerical diagonalization of the Bohr
  Hamiltonian in an $\grpso{5}$ basis.  Dynamical $\gamma$ deformation
  is found to significantly influence the predictions through its
  coupling to the rotational motion.  Basic signatures for the onset
  of rigid triaxial deformation are also obtained.
\end{abstract}

% 21.60.Ev collective models
% 21.10.Re collective levels

\maketitle
%***************************************************************************
% main text
%***************************************************************************

\section{Introduction}
\label{sec-intro}

The Bohr Hamiltonian~\cite{bohr1952:vibcoupling,bohr1998:v2}, together
with its
generalizations~\cite{kumar1966:bohr-solution,eisenberg1987:v1}, has
long served as the conceptual benchmark for interpreting quadrupole
collective dynamics in nuclei.  The conventional approach to numerical
diagonalization of the Bohr Hamiltonian, in a five-dimensional
oscillator
basis~\cite{gneuss1969:gcm,hess1980:gcm-details-238u,eisenberg1987:v1},
is slowly convergent and requires a large number of basis states to
describe a general deformed rotor-vibrator nucleus.  Therefore, it has
commonly been necessary to apply varying degrees of approximation in
addressing the dynamics of transitional and deformed nuclei, as in the
rotation-vibration model~\cite{faessler1965:rvm} and rigid triaxial
rotor~\cite{davydov1958:arm-intro} treatments of the Bohr Hamiltonian,
or in more recent studies of critical
phenomena~\cite{iachello2001:x5,iachello2003:y5,bonatsos2007:gamma-separable-x5,bonatsos2007:gamma-separable-davidson}.

However, diagonalization of the Bohr Hamiltonian is now
possible~\cite{rowe2009:acm} for potentials of essentially arbitrary
stiffness.  In particular, the algebraic collective model
(ACM)~\cite{rowe2004:tractable-collective,rowe2005:algebraic-collective,rowe2005:radial-me-su11,caprio2005:axialsep,rowe2010:rowanwood}
provides an efficient and straightforward computational framework
based on $\grpsutimes$ algebraic methods.  The Bohr Hamiltonian is
diagonalized in a basis of $\grpsutimes$ product wave functions
on the Bohr
deformation variables $\beta$ and $\gamma$ and Euler angles $\Omega$.
These are of the form $R_n^\lambda(a;\beta)\Psi_{v\alpha LM}(\gamma,\Omega)$,
where $R_n^\lambda$ is an $\grpsu{1,1}$ modified oscillator wave
function~\cite{rowe1998:davidson} and $\Psi_{v\alpha LM}$ is an
$\grpsochain$ spherical
harmonic~\cite{rowe2004:spherical-harmonics,caprio2009:gammaharmonic}.
The $\grpsochain$ formulation may be used either simply to extend the
conventional oscillator basis to higher phonon numbers sufficient to
provide full
convergence~\cite{debaerdemacker2007:so5-cartan,debaerdemacker2008:collective-cartan,debaerdemacker2009:acm-spectra}
or, further, to obtain much faster convergence as a function of
basis size through the use of
$\grpsu{1,1}$ $\beta$ wave functions chosen optimally for the nuclear
deformation~\cite{rowe2005:algebraic-collective}.

The Bohr Hamiltonian can consequently be applied, without
approximation, to the full range of nuclear quadrupole
rotational-vibrational structure, from spherical oscillator to axial
rotor to triaxial rotor.  Full convergence can be obtained for
energies and electromagnetic transition strengths involving high-lying
states, for instance, interband transitions among $\beta$, $\gamma$,
and multiphonon bands in well-deformed rotor nuclei.  The Bohr
Hamiltonian inherently induces coupling of the $\beta$, $\gamma$, and
rotational degrees of freedom, thereby yielding a rich set of
phenomena.

To approach an understanding of the full problem, we shall consider, in
this article, the simpler but already extensive implications of
coupling of the $\gamma$ and rotational degrees of freedom.  The
relevant Hamiltonian is then the ``angular'' part of the Bohr
Hamiltonian, and the ACM calculation reduces to diagonalization in a
basis of $\grpsochain$ spherical harmonics (Sec.~\ref{sec-method}).
The regime we address consists of rotational structure with axially
symmetric (axial) or weakly triaxial deformation.  However, even for a
nominally axial rotor, the Bohr description is found to mandate
significant dynamical fluctuations in $\gamma$, far from
$\gamma=0^\circ$.  The evolution of spectroscopic quantities (energies
and transition matrix elements) with respect to the $\gamma$
confinement provided by the potential is systematically investigated
(Sec.~\ref{sec-axial}), and the spectroscopic implications of the
onset of rigid triaxial structure are explored (Sec.~\ref{sec-triax}).
Probability distributions with respect to $\gamma$ and
with respect to the $K$ quantum number are then used to examine the
degree of adiabaticity, or separation of rotational and vibrational degrees of
freedom in the wave functions (Sec.~\ref{sec-decomp}).  Preliminary results
were presented in
Refs.~\cite{caprio2008:acm-cgs13,caprio2009:gammatriax}.

\section{Hamiltonian and solution method}
\label{sec-method}

\subsection{Hamiltonian}
\label{sec-method-hamiltonian}

The Bohr Hamiltonian~\cite{bohr1998:v2} is given, in terms of the
quadrupole deformation variables $\beta$ and $\gamma$ and Euler angles
$\Omega$, by
\begin{equation}
\label{eqn-Hbohr}
H=-\frac{\hbar^2}{2B}\biggl[
\frac{1}{\beta^4}\frac{\partial}{\partial\beta}\beta^4\frac{\partial}{\partial\beta}
-\frac{\Lambdahat^2}{\beta^2}\biggr]+V(\beta,\gamma),
\end{equation}
where
\begin{equation}
\label{eqn-Lambdasqr}
\Lambdahat^2=-\biggl(
\frac{1}{\sin 3\gamma} 
\frac{\partial}{\partial \gamma} \sin 3\gamma \frac{\partial}{\partial \gamma}
- \frac{1}{4}
\sum_\kappa \frac{\hat{L}_\kappa^{\prime2}}{\sin^2(\gamma -
\frac{2}{3} \pi \kappa
)}
\biggr).
\end{equation}
The operator appearing in brackets in the kinetic energy is the
Laplacian in five dimensions.  Its angular part $\Lambdahat^2$ is
the Casimir operator for the five-dimensional rotation group
$\grpso{5}$, which contains the rotations in physical space, acting on
the Euler angle coordinates, as an $\grpso{3}$ subgroup. 
The Bohr coordinates are five-dimensional spherical
polar coordinates, in terms of which the five components 
$q_M$ ($M=-2$, $\ldots$, $2$) of the quadrupole
deformation tensor are expressed as
\begin{multline}
\label{eqn-bohr-q}
q_M = \beta \biggl[ \cos\gamma\, \scrD^{(2)}_{0,M}(\Omega) 
\ifproofpre{\\+}{}
\frac{1}{\sqrt{2}} \sin\gamma \bigl[ \scrD^{(2)}_{2,M}(\Omega) +
\scrD^{(2)}_{-2,M}(\Omega) \bigr] \biggr].% 
\end{multline}
The potential energy
$V(\beta,\gamma)$ must be periodic in $\gamma$, with period
$120^\circ$, and it must be symmetric about $\gamma=0^\circ$ and
$\gamma=60^\circ$.  
The Bohr coordinate system and Hamiltonian are
reviewed in detail in, \textit{e.g.},
Ref.~\cite{prochniak2009:bohr-collective-hf-hg}.  

The restriction to angular coordinates $(\gamma,\Omega)$ then yields a
Hamiltonian 
\begin{equation}
\label{eqn-gammarotor}
H=\Lambdahat^2+V(\gamma).
\end{equation}
Such an angular Hamiltonian arises as a schematic limit of the full
Bohr Hamiltonian when the coordinate $\beta$ in~(\ref{eqn-Hbohr}) is
taken to be rigidly fixed, as might be considered for a well-deformed
nucleus.  However, a reduction to the angular
form~(\ref{eqn-gammarotor}) is more broadly applicable to transitional
nuclei as
well~\cite{bonatsos2007:gamma-separable-x5,bonatsos2007:gamma-separable-davidson},
since it occurs by separation of variables when the potential is of
the form
$V(\beta,\gamma)=u(\beta)+v(\gamma)/\beta^2$~\cite{jean1960:transition-model}.
The explicit relations for reduction to an angular
Hamiltonian are
reviewed in Appendix~\ref{app-angular}.
The symmetry conditions on $V(\gamma)$ are satisfied by the function
$\cos3\gamma$ and powers $\cos^n 3\gamma$ thereof.
%----------------------------------------------------------------
\begin{figure}
\begin{center}
\includegraphics*[width=\ifproofpre{1}{0.6}\hsize]{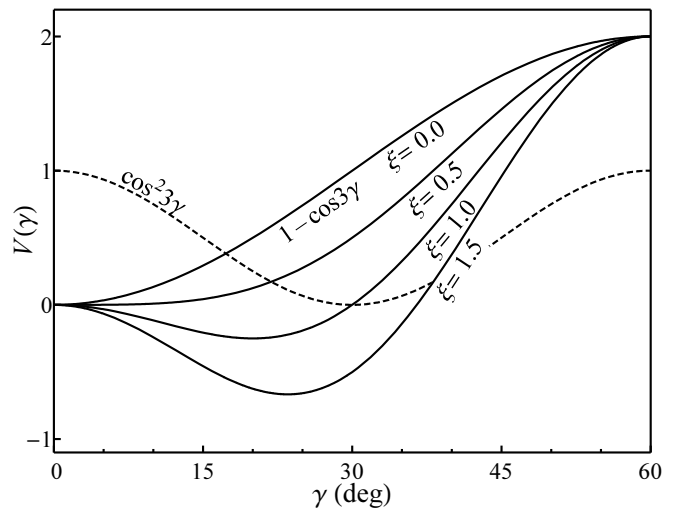}
\end{center}
\caption{The shape of the potential $V(\gamma)$ used
  in~(\ref{eqn-gammarotor-xi}), plotted for
  various values of $\xi$ (taking $\chi=1$).  Note that a constant
  offset $\xi$ has been  subtracted from each curve, so that
  $V(0)=0$ in each case.  The dotted curve indicates the shape of the
  contribution from  $\cos^2 3\gamma$.}
\label{fig-gammarotor-potl}
\end{figure}
%----------------------------------------------------------------

Let us therefore consider, in particular,   
\begin{equation}
  \label{eqn-gammarotor-xi}
H=\Lambdahat^2 + \chi \bigl[ (1-\cos 3\gamma) + \xi \cos^2 3\gamma\bigr].
\end{equation}
The possible shapes of the potential appearing in this Hamiltonian are
shown in Fig.~\ref{fig-gammarotor-potl}.  For $\xi=0$,
$V(\gamma)\propto(1-\cos3\gamma)$, as considered in
Ref.~\cite{rowe2004:tractable-collective}, providing a minimum at
$\gamma=0^\circ$ (axial deformation).  With increasing $\chi$, a
``deeper'' potential provides greater confinement or stabilization
around $\gamma=0^\circ$, approximately harmonic ($\propto \gamma^2$)
for small $\gamma$.  Including a $\cos^2 3\gamma$ term
[Fig.~\ref{fig-gammarotor-potl} (dotted curve)] by taking $\xi$
nonzero introduces a richer extremum structure and a means for
studying the axial-triaxial shape transition~\cite{iachello2003:y5}.
For $\xi=1/2$, the potential is more softly confining in $\gamma$,
with a quartic minimum (locally $\propto \gamma^4$).  This case is
termed ``critical'' in Ref.~\cite{iachello2003:y5}.  For $\xi>1/2$,
the potential has a minimum at a nonzero value of $\gamma$, given by
$\cos3\gamma_0=1/(2\xi)$.  For large positive $\xi$, the $\cos^2
3\gamma$ term dominates, and the minimum approaches $\gamma=30^\circ$.
Although not considered here, with a negative $\cos^2 3\gamma$
contribution the Hamiltonian~(\ref{eqn-gammarotor-xi}) may also be
used to investigate prolate-oblate shape
coexistence~\cite{sato2010:triax-dynamics}.

\subsection{Solution method}
\label{sec-method-soln}

Any function of the coordinates $(\gamma,\Omega)$ with the requisite
symmetry properties for a wave function can be
expressed in terms of symmetric linear combinations of Wigner $\scrD$
functions as (\textit{e.g.}, Ref.~\cite{prochniak2009:bohr-collective-hf-hg}) 
\begin{equation}
\label{eqn-expansion-F}
\psi(\gamma,\Omega)=\sum_{\substack{K=0\\\text{even}}}^L
F_{K}(\gamma) \xi^{(L)}_{KM}(\Omega),
\end{equation}
where~\cite{caprio2009:gammaharmonic}
\begin{equation} 
\label{eqn-xi}
\xi^{(L)}_{KM}(\Omega) \equiv \frac{1}{\deltasqrt{K}}
\Bigl[\scrD^{(L)}_{KM}(\Omega)+(-)^L\scrD^{(L)}_{-KM}(\Omega)\Bigr].
\end{equation} 
The wave function is thus fully specified by the $F_K(\gamma)$.  

A complete set for expanding wave functions $\psi(\gamma,\Omega)$ is provided by
the $\grpsochain$ spherical harmonics $\Psi_{v\alpha
  LM}(\gamma,\Omega)$~\cite{rowe2004:spherical-harmonics,caprio2009:gammaharmonic}.
The $\grpsochain$ spherical harmonics are defined as the
eigenfunctions of the $\grpso{5}$ Casimir operator $\Lambdahat^2$, with
\begin{equation}
\label{eqn-harmonic-eigen}
\Lambdahat^2 \Psi_{v\alpha{}LM}  (\gamma,\Omega) = v (v+3)
\Psi_{v\alpha{}LM} (\gamma,\Omega),
\end{equation}
chosen furthermore to posess definite angular momentum with respect to
the $\grpso{3}$ subgroup of physical rotations.  The $\Psi_{v\alpha
  LM}$ are labeled by the $\grpso{5}$ seniority quantum number $v$
($v=0$, $1$, $\ldots$), the $\grpso{3}$ angular momentum quantum
number $L$, and its $z$-projection quantum number $M$.  (A
multiplicity index $\alpha$ is also required to complete the labeling
for $v\geq 6$ but will be omitted from the notation below when not
needed.)  The $\Psi_{v\alpha LM}$ are explicitly realized by
constructing the functions $F_K(\gamma)$ needed to express each
spherical harmonic in the form~(\ref{eqn-expansion-F}), as may be
accomplished by the algorithm of
Refs.~\cite{rowe2004:spherical-harmonics,caprio2009:gammaharmonic}.

Diagonalization of the Hamiltonian~(\ref{eqn-gammarotor-xi}) is
carried out in a finite basis of these $\grpsochain$ spherical
harmonics, truncated to some maximum seniority $\vmax$.  In general,
higher-seniority spherical harmonics are needed for the construction
of more highly $\gamma$-localized wave functions.  Thus,
diagonalization for Hamiltonians with stiffer $\gamma$ confinement requires a basis with
higher $\vmax$.  A basis with $\vmax= 50$ amply suffices for
convergence of all calculations in the present work.

It is first necessary to compute the Hamiltonian matrix elements with
respect to the $\grpsochain$ basis.  For the kinetic energy, the
matrix elements
$\tme{\Psi_{v'\alpha'{}LM}}{\Lambdahat^2}{\Psi_{v\alpha{}LM}}$ are
trivially evaluated by the eigenvalue
equation~(\ref{eqn-harmonic-eigen}).  For the potential energy, the
matrix elements of $\cos 3\gamma$ may be evaluated in terms of
integrals of products of $F_K(\gamma)$
functions~\cite{rowe2004:tractable-collective}.  Since
$\Psi_{300}(\gamma,\Omega)=(8\pi^2)^{-1/2} (3/\sqrt{2})\cos 3\gamma$,
it may be noted that the matrix elements of interest are triple
overlaps $\tme{\Psi_{v'\alpha'{}LM}}{\Psi_{300}}{\Psi_{v\alpha{}LM}}$
of spherical harmonics, which are equivalent to $\grpsochain$
generalized Clebsch-Gordan
coefficients~\cite{rowe2004:spherical-harmonics,caprio2009:gammaharmonic}.
These are calculated and tabulated electronically (for $v\leq50$) in
Ref.~\cite{caprio2009:gammaharmonic}.  The matrix elements of
$\cos^n3\gamma$ follow immediately from those of $\cos 3\gamma$, by
insertion of resolutions of the identity, \textit{i.e.}, by matrix
multiplication.

Then, diagonalization of the Hamiltonian matrix yields the
amplitudes $a_{Lij}$ in the decomposition
\begin{equation}
\label{eqn-diag-expansion}
\psi_{LiM}(\gamma,\Omega)=\sum_j a_{Lij}\Psi_{LjM}(\gamma,\Omega).
\end{equation}  Here we have denoted the $i$th eigenfunction of the
Hamiltonian, for angular
momentum $L$, by $\psi_{LiM}(\gamma,\Omega)$ and likewise relabeled 
the $j$th $\grpsochain$ spherical harmonic of angular momentum $L$ as $\Psi_{LjM}$, 
\textit{i.e.}, replacing $v$ and $\alpha$ by a simple running index~\cite{caprio2009:gammaharmonic}.  

The leading-order electric quadrupole operator in the Bohr framework
is $\calM(E2)\propto q$.  Under the present restriction to angular
coordinates, $\calM(E2)\propto \calQ$, where $\calQ$ is the
\textit{unit} quadrupole tensor~\cite{rowe2004:spherical-harmonics},
defined by $q_M=\beta\calQ_M$ [see~(\ref{eqn-bohr-q})].  It is
straightforward to calculate transition matrix elements between the
Hamiltonian eigenstates~(\ref{eqn-diag-expansion}), once the matrix
elements are obtained between the basis states.  Since
$\Psi_{12M}(\gamma,\Omega)=(8\pi^2)^{-1/2} \sqrt{15/2}\calQ_M$, the
reduced matrix elements are proportional to
$\trme{\Psi_{v'\alpha'{}L'}}{\Psi_{12}}{\Psi_{v\alpha{}L}}$, which are
again given by $\grpsochain$ generalized Clebsch-Gordan coefficients,
available from Ref.~\cite{caprio2009:gammaharmonic}.

\section{Phonon and multiphonon excitations 
}
\label{sec-axial}

\subsection{Spectra}
\label{sec-axial-spectra}
%----------------------------------------------------------------
\begin{figure*}[p]
\begin{center}
\includegraphics*[width=0.95\hsize]{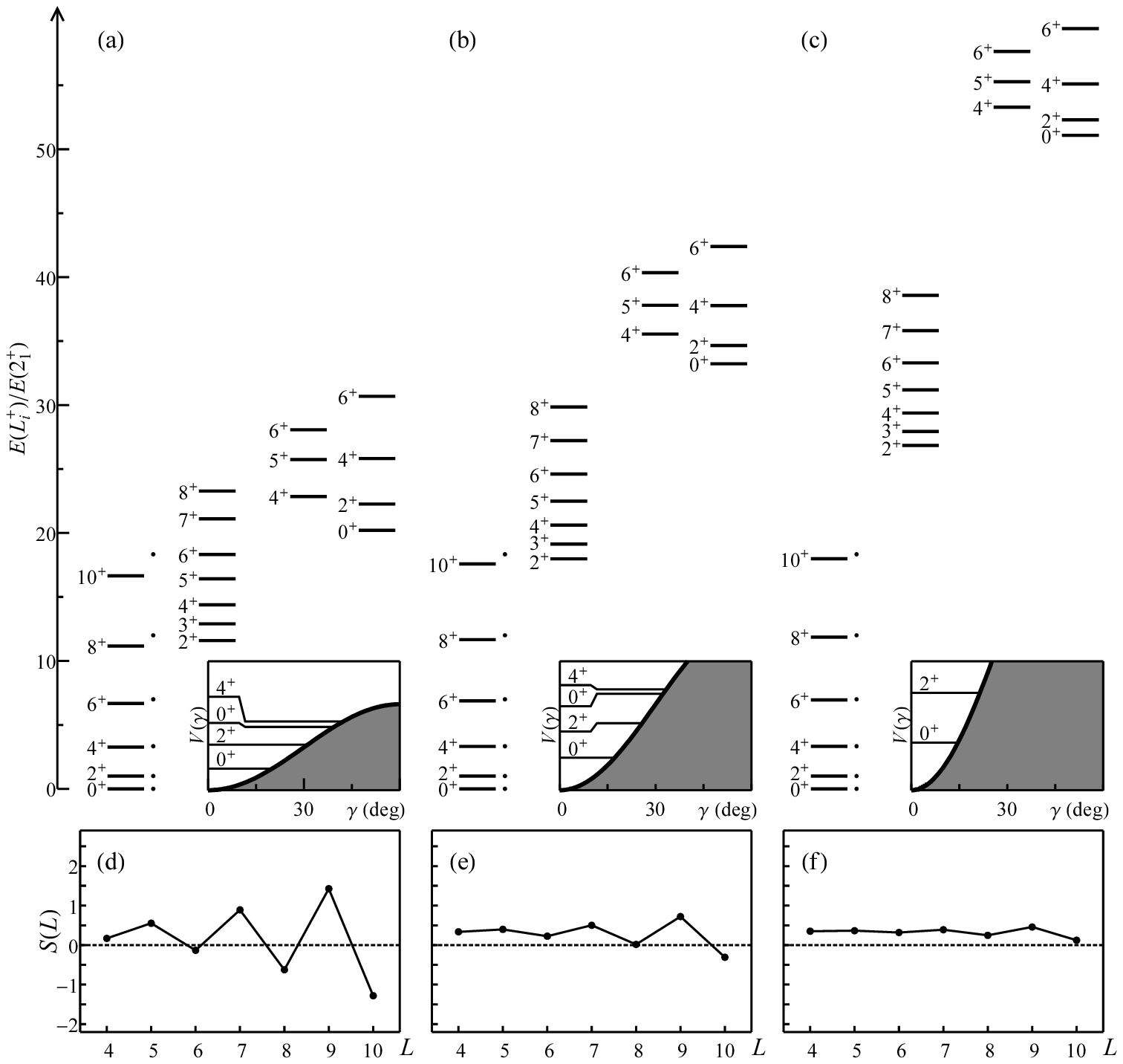}
\end{center}
\caption{
Level schemes for the angular Hamiltonian~(\ref{eqn-gammarotor-xi})
with $\xi=0$, for (a)~$\chi=50$, (b)~$\chi=100$, and (c)~$\chi=200$.
Rotational $L(L+1)$ energies for the yrast band are indicated by the
dots. The potential $V(\gamma)$ is shown in the inset, with the
ground, quasi-$\gamma$, and quasi-$\gamma\gamma$ bandhead energies
indicated.  (d--f)~Staggering of level energies within the
quasi-$\gamma$ band, as measured by the energy second difference
$S(L)$.  }
\label{fig-gammarotor-axial-schemes}
\end{figure*}
%----------------------------------------------------------------
%----------------------------------------------------------------
\begin{figure}
\begin{center}
\includegraphics*[width=\ifproofpre{0.95}{0.6}\hsize]{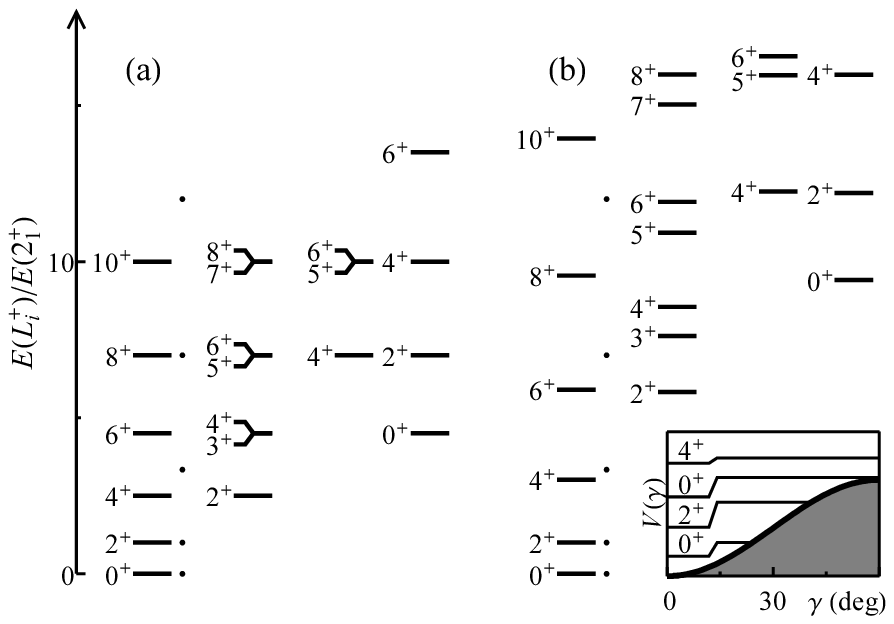}
\end{center}
\caption{ Level schemes for the angular
  Hamiltonian~(\ref{eqn-gammarotor-xi}) with $\xi=0$, for (a)~the
  $\gamma$-independent limit $\chi=0$ and (b)~$\chi=20$, with levels
  arranged anticipating the quasiband structure of
  Fig.~\ref{fig-gammarotor-axial-schemes}.  Rotational $L(L+1)$
  energies for the yrast band are indicated by the dots. The potential
  $V(\gamma)$ for $\chi=20$ is shown in the inset, with the ground,
  quasi-$\gamma$, and quasi-$\gamma\gamma$ bandhead energies
  indicated.  }
\label{fig-gammarotor-axial-schemes-soft}
\end{figure}
%----------------------------------------------------------------

The nature of the spectra obtained from the
Hamiltonian~(\ref{eqn-gammarotor-xi}) depends both on the
\textit{depth} of the potential (determined by $\chi$) and the
\textit{shape} of the potential (determined by $\xi$ as in
Fig.~\ref{fig-gammarotor-potl}).  The depth of the potential
effectively controls the degree of $\gamma$ confinement.  It is worth
first carefully considering the implications of $\gamma$ confinement,
or conversely $\gamma$ softness, within this Bohr Hamiltonian
framework.  In this section, we shall therefore investigate the
structural dependence on $\chi$ (for $\xi=0$), before proceeding to
the dependence of structure on the shape of the potential, and in
particular the onset of rigid triaxiality, in Sec.~\ref{sec-triax}.

The results of illustrative calculations are shown in
Fig.~\ref{fig-gammarotor-axial-schemes}, for $\chi=50$, $100$, and
$200$.    The low-lying states form
quasi-bands which may be roughly identified as a ground-state
rotational band ($K=0$), $\gamma$ vibrational excitation ($K=2$), and
two-phonon $\gamma$ excitations ($K=4$ and $0$), denoted by
$\gamma\gamma_4$ and $\gamma\gamma_0$.

The stiffness of the potential around $\gamma=0^\circ$ simultaneously
determines both the $\gamma$-vibrational energy scale [increasing from
Fig.~\ref{fig-gammarotor-axial-schemes}(a)
to Fig.~\ref{fig-gammarotor-axial-schemes}(c)] and also how well confined
the wave function is with respect to $\gamma$, as seen in the
corresponding approach to an ideal rotational spectrum.  Thus, within
the framework of the Bohr Hamiltonian, the $\gamma$ band energy~--- more
specifically, the energy ratio $E(2^+_\gamma)/E(2^+_1)$, or separation
of vibrational and rotational energy scales~--- and the $\gamma$ softness
of the wave function are inextricably linked. 

As a starting point, it may be observed that for $\chi=0$ the
potential is strictly $\gamma$-independent, and the spectrum therefore
follows
an $\grpso{5}$ multiplet
structure~\cite{wilets1956:oscillations,rakavy1957:gsoft}.  Successive
multiplets consist of angular momenta $0$, $2$, $4$-$2$, $6$-$4$-$3$-$0$,
$\ldots$, for $v=0$, $1$, $2$, $3$, $\ldots$, respectively, with multiplet
energies $\propto v(v+3)$, as depicted in
Fig.~\ref{fig-gammarotor-axial-schemes-soft}(a).  The system is simply
a Wilets-Jean~\cite{wilets1956:oscillations} or
$\grpso{6}$~\cite{arima1979:ibm-o6} rotor, but without $\beta$
excitations (see also
Ref.~\cite{rowe2004:tractable-collective}).  Then, as $\gamma$ confinement is introduced, the
familiar rotational band structure begins to emerge.  An intermediate
spectrum, obtained for $\chi=20$, is shown in
Fig.~\ref{fig-gammarotor-axial-schemes-soft}(b).

For $\chi=50$
[Fig.~\ref{fig-gammarotor-axial-schemes}(a)], rotational quasi-bands
are well-developed, and
$E(2^+_\gamma)/E(2^+_1)\approx10$, as appropriate to, \textit{e.g.},
the well-deformed rare earth nuclei.  However, it is seen from the
potential plot in Fig.~\ref{fig-gammarotor-axial-schemes}(a) that the
$\gamma$ confinement for this value of $\chi$ is still weak.  The
range of energetically accessible $\gamma$ values increases
significantly for successive phonon excitations, such that confinement
is almost nonexistent at the energy of the two-phonon excitation.

Dynamical $\gamma$ deformation consequently plays a major role in the
calculated structure, through its interaction with the rotational
dynamics.  This is reflected in significant deviations from ideal
rotational behavior in the spectroscopic predictions.

Most noticeably, on inspection of
Fig.~\ref{fig-gammarotor-axial-schemes}(a), level energies within the
$\gamma$ quasi-band follow a gently $\gamma$-soft staggering pattern
[$2(34)(56)\ldots$].  This staggering is reminiscent of the
$\grpso{5}$ level degeneracies obtained for $\chi=0$, and it disappears as
the $\gamma$ stiffness increases
[Fig.~\ref{fig-gammarotor-axial-schemes}(b,c)].  
The deviations from rotational energy spacings are even more
pronounced for the calculated two-phonon bands.  Note
especially the near doubling of the rotational energy spacing scale
for the two-phonon bands, relative to the ground state band, for
$\chi=50$ [Fig.~\ref{fig-gammarotor-axial-schemes}(a)].  

The deviations from rotational energy spacings within the $\gamma$
band may
be seen most clearly from plots of the level energy second difference
$S(L)\equiv\bigl[[E(L)-E(L-1)]-[E(L-1)-E(L-2)]\bigr]/E(2^+_1)$, as
shown in Fig.~\ref{fig-gammarotor-axial-schemes}(d--f).  For an
ideal rotational band with $L(L+1)$ energy spacings, the curve is
flat, with $S(L)=1/3$.  Alternatively, $\gamma$-soft staggering is
manifest in minima at even $L$.  As surveyed in
Ref.~\cite{mccutchan2007:gamma-staggering}, the observed level
energies within the $\gamma$ bands of most transitional and rotational
nuclei yield $S(L)$ plots which are either gently $\gamma$-soft or
near constant ($\approx1/3$).  A few transitional
nuclei (\textit{e.g.},
 $\isotope[152]{Sm}$, $\isotope[156]{Gd}$, or $\isotope[162]{Er}$) exhibit a degree of
staggering comparable to that found for $\chi=50$ (see also Refs.~\cite{caprio2005:axialsep,dusling2006:160er-fusevap}).
However, most rare earth rotational nuclei (see Fig.~3 of
Ref.~\cite{mccutchan2007:gamma-staggering}) more clearly follow an
$L(L+1)$ energy spacing within the $\gamma$ band.  There is thus an
apparent
 disagreement between the degree of dynamical $\gamma$
softness expected in the Bohr picture given
$E(2^+_\gamma)/E(2^+_1)\approx10$, and the observed structure in
nuclei, at least if we assume the basic
Hamiltonian~(\ref{eqn-gammarotor-xi}).  

Within the ground state band, the Hamiltonian~(\ref{eqn-gammarotor-xi})
is found to yield relative energies [\textit{i.e.},
$E(L^+_1)/E(2^+_1)$] which fall below the $L(L+1)$ expectation for an
adiabatic rotor.  The ideal rotational energies are indicated, for
comparison, by the dots in Fig.~\ref{fig-gammarotor-axial-schemes}(a--c).
The deviation from $L(L+1)$ spacing within the ground
state band decreases, as would be expected, for increasing $\gamma$
stiffness.
The effect has already been noted in the context of a full $\beta$ and
$\gamma$ calculation with the ACM in Ref.~\cite{rowe2009:acm} (see
Fig.~5 of that reference).
Such a deviation would traditionally be characterized as ``centrifugal stretching'',
based on an the interpretation in which the $\beta$
deformation increases, and thus the rotational moments increase, with
increasing angular momentum.  However, here the effect is seen to arise purely
from the interaction of $\gamma$ and rotational degrees of freedom,
for a system in which ``stretching'' in the $\beta$ degree of freedom is strictly
impossible.  

\subsection{Evolution of observables}
\label{sec-axial-evoln}
%----------------------------------------------------------------
\begin{figure*}
\begin{center}
\includegraphics*[width=0.9\hsize]{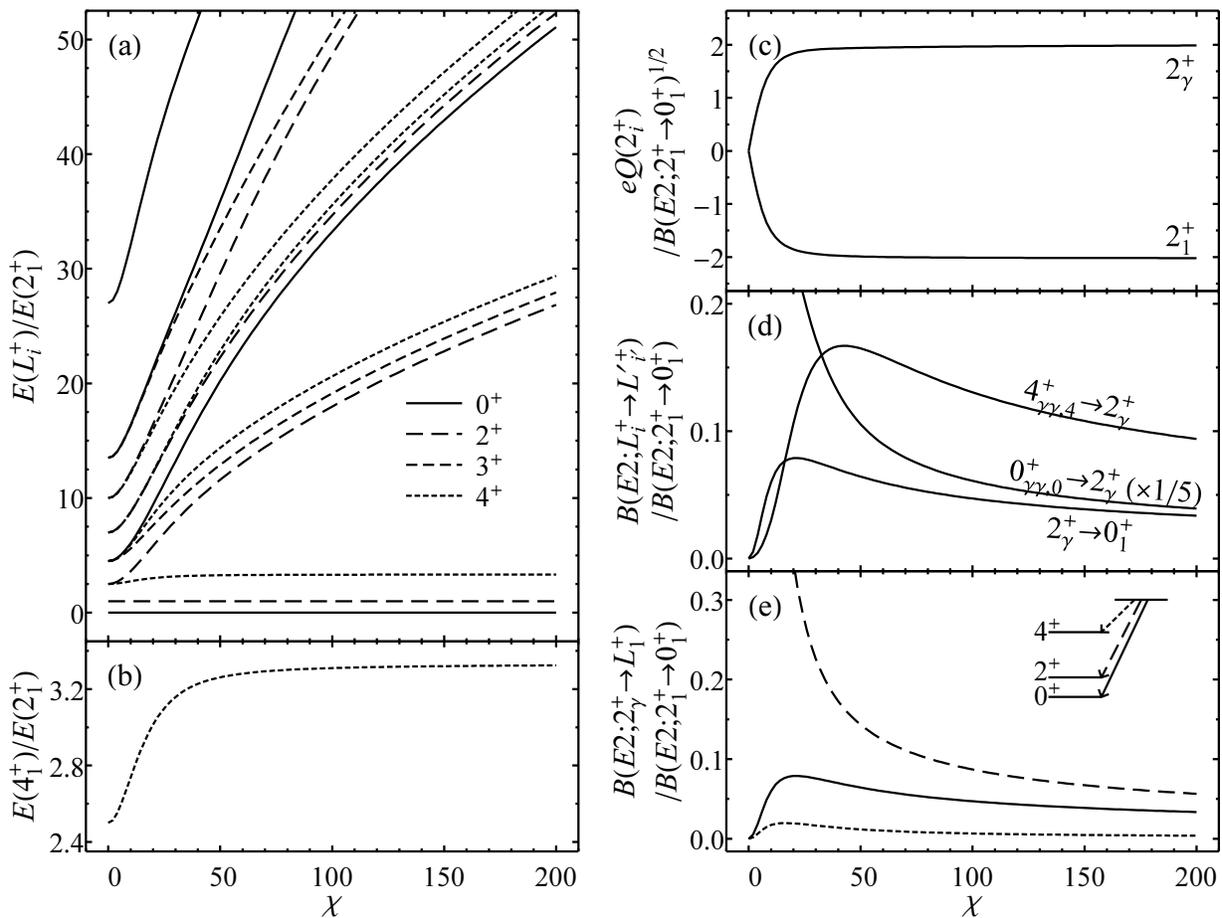}
\end{center}
\caption{Evolution of spectroscopic properties with $\gamma$ stiffness,
  for the angular Hamiltonian~(\ref{eqn-gammarotor-xi}) with $\xi=0$.  Quantities shown are (a)~excitation
  energies of low-lying levels, normalized to $E(2^+_1)$, (b)~the
  energy ratio $E(4^+_1)/E(2^+_1)$, specifically, (c)~electric
  quadrupole moments of the ground state band and $\gamma$ band $2^+$
  members, (d)~electric quadrupole reduced transition probabilities
  for one-phonon transitions between the ground, $\gamma$, and
  two-phonon $\gamma$ ($K=0$ and $4$) bandhead states, and (e)~reduced
  transition probabilities for the transitions depopulating the
  $2^+_\gamma$ bandhead state.  All electromagnetic quantities are normalized
  to $B(E2;2^+_1\rightarrow0^+_1)\equiv1$.
}
\label{fig-gammarotor-evoln}
\end{figure*}
%----------------------------------------------------------------

The evolution of the numerical predictions, with increasing $\gamma$
stiffness, is examined more quantitatively and systematically in
Fig.~\ref{fig-gammarotor-evoln}.  Both the energy spectrum
[Fig.~\ref{fig-gammarotor-evoln}(left)] and electromagnetic
(specifically, electric quadrupole) moments and transition matrix
elements [Fig.~\ref{fig-gammarotor-evoln}(right)] are shown, as functions of $\chi$.

The onset and evolution of rotational band structure, as $\gamma$
confinement is introduced, may be traced in the full energy spectrum
[Fig.~\ref{fig-gammarotor-evoln}(a)].  Note especially the correlation
between the $\gamma$ band energy [Fig.~\ref{fig-gammarotor-evoln}(a)]
and the ground state band energy ratio $E(4^+_1)/E(2^+_1)$
[Fig.~\ref{fig-gammarotor-evoln}(b)], which varies from $2.5$ for
$\gamma$-independent rotation to $3.33$ for rigid axial rotation.
This ratio is commonly taken as an indicator of rotational
adiabaticity.  For the present restricted problem, adiabaticity
represents separation of the $\gamma$ and rotational degrees of
freedom, but in general for the Bohr Hamiltonian the quantitative
details will also be affected by the $\beta$ degree of freedom.  The
evolution of multiphonon band energies can also be followed in
Fig.~\ref{fig-gammarotor-evoln}.  These begin anharmonically low, at
less than twice the $\gamma$ band energy~---
for $\chi=50$, an estimate based on low-lying band members gives
$E_{\gamma\gamma,4}/E_\gamma\approx1.7$ and
$E_{\gamma\gamma,0}/E_\gamma\approx1.9$~--- but approach harmonicity
as $\chi$ increases.  The relative energies of the bands may also be
seen in Fig.~\ref{fig-gammarotor-axial-schemes}(a--c).

The evolution of electromagnetic properties is traced for
representative quadrupole moments and transition strengths in
Fig.~\ref{fig-gammarotor-evoln}(right).  In the $\gamma$-independent
limit, the wave functions are simply the $\grpsochain$ spherical
harmonics themselves, and electromagnetic matrix elements are governed by
$\grpso{5}$ selection rules and related by $\grpsochain$ Clebsch-Gordan
coefficients.  On the other hand, in the limit of large $\gamma$
stiffness, electromagnetic matrix elements are expected to approach
the Alaga rule ratios~\cite{alaga1955:branching,bohr1998:v2} of the
adiabatic axial rotor, given by ordinary angular momentum
Clebsch-Gordan coefficients.

The electric quadrupole moments $Q(2^+_1)$ and $Q(2^+_\gamma)$ are
shown in Fig.~\ref{fig-gammarotor-evoln}(c).  All quadrupole moments
vanish in the $\gamma$-independent limit, by a selection rule arising
from a parity quantum number defined in the five-dimensional space of
the Bohr coordinates
($\bbR^5$-parity)~\cite{bes1959:gamma,rowe2009:acm,caprio2009:gammaharmonic}.
In the rotational limit, these quadrupole moments are expected to
approach values of $\pm8\sqrt{\pi}/7\approx\pm2.03$, negative for the
ground state band ($K=0$) and positive for the $\gamma$ band ($K=2$),
expressed relative to $B(E2;2^+_1\rightarrow0^+_1)^{1/2}$.  These
values are rapidly attained, by $\chi\lesssim25$.

For harmonic $\gamma$ vibration, the $\gamma\rightarrow g$,
$\gamma\gamma_4\rightarrow \gamma$, and
$\gamma\gamma_0\rightarrow\gamma$ interband intrinsic matrix elements
$\tme{f}{\calM'}{i}$~\cite{bohr1998:v2} are expected to be in the
proportion $1:\sqrt{2}:1$~\cite{rowe2010:rowanwood}.  The overall
normalization of these intrinsic matrix elements, \textit{i.e.}, the
$\gamma\rightarrow{}g$ strength, decreases with increasing $\gamma$
stiffness~\cite{eisenberg1987:v1}.  For the transitions among the
bandhead states, in particular, these intrinsic matrix element ratios
correspond to $B(E2;2^+_\gamma\rightarrow 0^+_g)$,
$B(E2;4^+_{\gamma\gamma,4}\rightarrow 2^+_\gamma)$, and
$B(E2;0^+_{\gamma\gamma,0}\rightarrow 2^+_\gamma)$ strengths in the
proportion $1:2.8:5$.  The approach to harmonic values is seen in
Fig.~\ref{fig-gammarotor-evoln}(d). Simply from considering these
transitions, harmonic behavior would appear to set in very gradually
for $\chi\gtrsim 50$.  However, a more comprehensive consideration of
the electromagnetic transition strengths, which leads to some
modification of this conclusion, is provided by the Mikhailov analysis
in Sec.~\ref{sec-axial-mikhailov}.  The branching ratios for electric
quadrupole transitions between bands likewise approach the Alaga rule
ratios.  For the transitions from the $2^+_\gamma$ bandhead to the
ground state band members [Fig.~\ref{fig-gammarotor-evoln}(e)], for
instance, the adiabatic rotor has $B(E2;2^+_\gamma\rightarrow0^+_g)$,
$B(E2;2^+_\gamma\rightarrow2^+_g)$ and
$B(E2;2^+_\gamma\rightarrow4^+_g)$ strengths in the proportion
$0.4:0.57:0.029$.

\subsection{\boldmath Effective $\gamma$ deformation}
\label{sec-axial-gamma}
%----------------------------------------------------------------
\begin{figure}
\begin{center}
\includegraphics*[width=\ifproofpre{1}{0.6}\hsize]{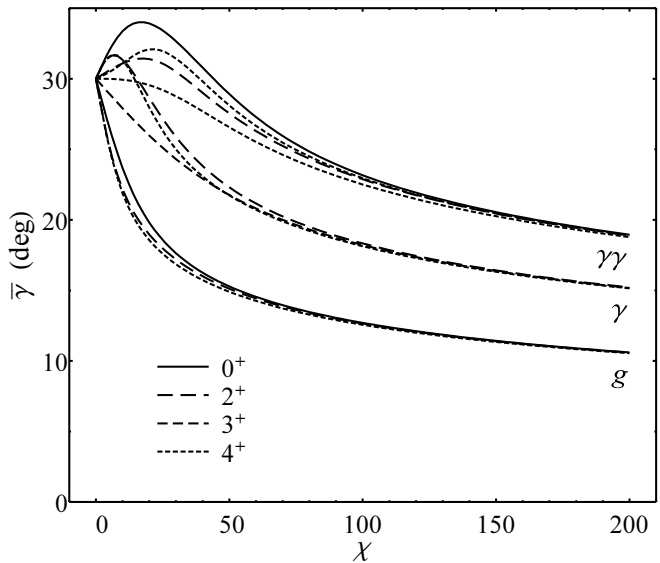}
\end{center}
\caption{
Evolution of the effective values $\gammabar$ with respect to
stiffness parameter $\chi$, for the angular
Hamiltonian~(\ref{eqn-gammarotor-xi}) with $\xi=0$.  Values are shown
for ground state, $\gamma$, $\gamma\gamma_4$, and $\gamma\gamma_0$
quasi-band members with $L\leq 4$. }
\label{fig-gammarotor-axial-gamma}
\end{figure}
%----------------------------------------------------------------

Although we have so far examined $\gamma$ softness indirectly, through
its spectroscopic signatures, the wave function $\psi(\gamma,\Omega)$
is directly accessible for the eigenstates calculated in the
diagonalization of the Bohr Hamiltonian, and thus the deviation of
$\gamma$ from $0^\circ$ can be considered directly.  The simplest
measure is provided by an effective $\gamma$
value $\gammabar$, defined by
\begin{equation}
\label{eqn-gammabar-defn}
\cos
3\gammabar \equiv\tbracket{\cos 3\gamma}.
\end{equation} 
The matrix elements of $\cos
3\gamma$ in the $\grpsochain$ spherical harmonic basis are already
available, as noted
in Sec.~\ref{sec-method}, so this expectation value may readily be
calculated.
The definition~(\ref{eqn-gammabar-defn}) is consistent with the quadrupole
shape invariant approach~\cite{kumar1972:q-invariant,cline1986:coulex}, in which 
an effective $\gamma$ for the full $(\beta,\gamma,\Omega)$ coordinate space is defined by
$\cos3\gamma_\text{eff}=\tbracket{\beta^3
  \cos3\gamma}/\tbracket{\beta^2}^{3/2}$~\cite{elliott1986:ibm-shape,jolos1997:ibm-shape-invariant,werner2005:triax-invariant}. 

The evolution of $\gammabar$ for the ground state, $\gamma$, and
$\gamma\gamma$ band members (for $L\leq4$) is shown in
Fig.~\ref{fig-gammarotor-axial-gamma}.  In the $\chi=0$
($\gamma$-independent) limit, $\tbracket{\cos 3\gamma}=0$ by the
$\bbR^5$-parity selection rule, and thus $\gammabar=30^\circ$ for
\textit{all} states.  As $\chi$ increases past $\chi\approx50$, it is
seen that the $\gammabar$ values for the members of each band cluster
and decrease with increasing $\chi$.  The $\gammabar$ value jumps
substantially between bands, increasing from ground to $\gamma$ to
$\gamma\gamma$ bands, indeed, as expected for successive phonon
excitations.

The situation for ``axial rotor'' nuclei within the Bohr Hamiltonian
framework is very much contrary to the classic but schematic
characterization of such nuclei as having ``$\gamma\approx0^\circ$'',
which may be more concretely interpreted as $\gamma\ll30^\circ$.  Recall
that the $\gamma$-band excitation energies matching the experimental
values for rotor nuclei are obtained for $\chi\approx50$.  For this
stiffness, the ground state band members have $\gammabar\approx15^\circ$, and the
$\gamma$ band members have $\gammabar\approx23^\circ$.  These large
$\gammabar$ values are consistent with the large range of
energetically accessible $\gamma$ values for these states
[Fig.~\ref{fig-gammarotor-axial-schemes}(a,inset)].  The full
probability distribution with respect to the $\gamma$ coordinate is
considered in Sec.~\ref{sec-decomp}.

\subsection{Intrinsic matrix elements}
\label{sec-axial-mikhailov}  
%----------------------------------------------------------------
\begin{figure*}
\begin{center}
\includegraphics*[width=0.7\hsize]{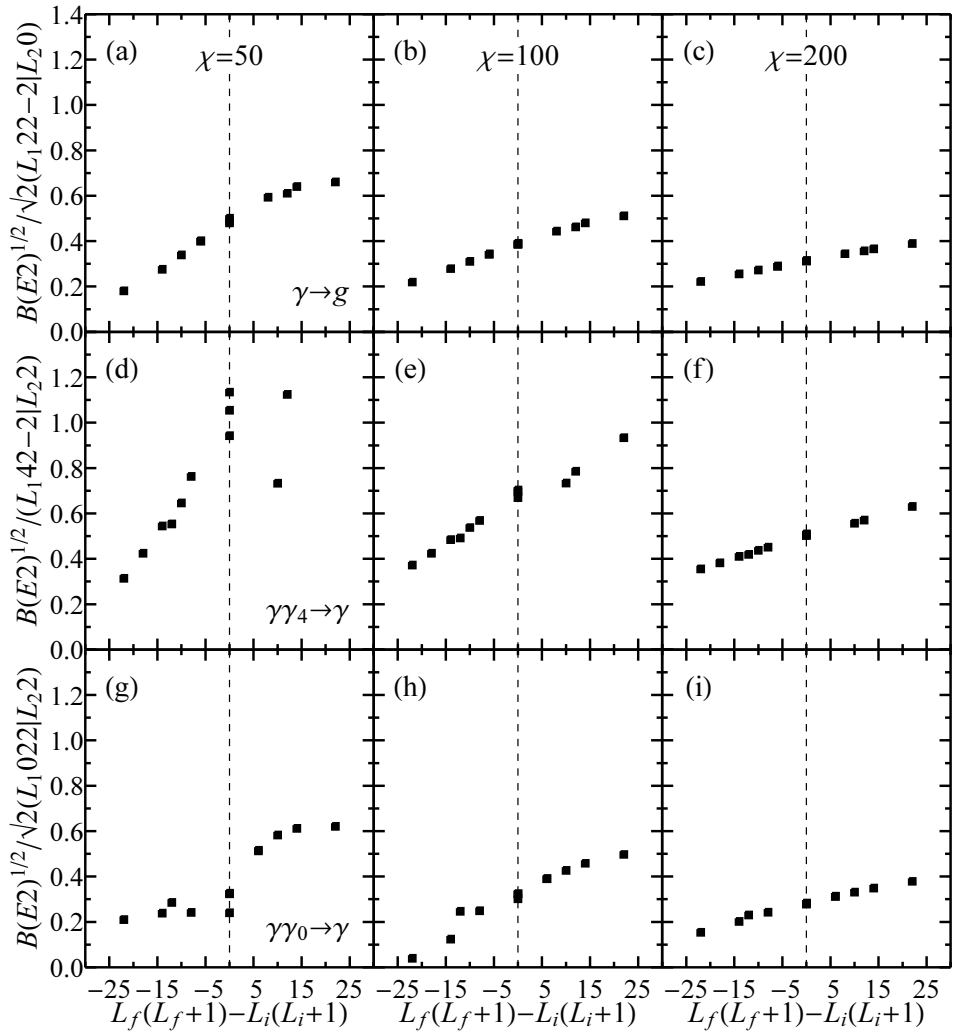}
\end{center}
\caption{
Interband transition amplitudes $B(E2)^{1/2}$, from the $\gamma$ quasi-band to the
ground state band~(top), from the $\gamma\gamma_4$ quasi-band
to the $\gamma$ quasi-band~(middle), and from the $\gamma\gamma_0$ quasi-band
to the $\gamma$ quasi-band~(bottom), for Mikhailov analysis.  Plots
are included for the calculations of Fig.~\ref{fig-gammarotor-axial-schemes}, with
$\chi=50$~(left), $\chi=100$~(middle),
and $\chi=200$~(right) and $\xi=0$.  The values shown are for
transitions between levels with $L\leq6$, normalized to
$B(E2;2^+_1\rightarrow0^+_1)\equiv1$.
}
\label{fig-gammarotor-axial-mikhailov}
\end{figure*}
%----------------------------------------------------------------
%----------------------------------------------------------------
%----------------------------------------------------------------
\begin{table*}
  \caption{Electric quadrupole interband intrinsic matrix elements $\tme{f}{\calM'}{i}$ 
    and mixing parameters $a$, for different
    $\gamma$ stiffnesses, as extracted from
    the Mikhailov analyses of Fig.~\ref{fig-gammarotor-axial-mikhailov}.
    Ratios, as indicators of anharmonicity, are tabulated in the final two
    columns.  The values
    for an adiabatic rotor with harmonic $\gamma$ vibration~\cite{rowe2010:rowanwood}  are
    included for comparison.  The values
    for the intrinsic matrix elements are normalized to 
    $B(E2;2^+_1\rightarrow0^+_1)\equiv1$.
  }
\label{tab-ime-axial}
\begin{center}
\begin{ruledtabular}
\begin{tabular}{ldddddddd}
\ifproofpre{\\[-8pt]}{\\[-24pt]} % for fractions
 &
\multicolumn{2}{c}{$\gamma\rightarrow g$}
&
\multicolumn{2}{c}{$\gamma\gamma_4\rightarrow\gamma$}
&
\multicolumn{2}{c}{$\gamma\gamma_0\rightarrow\gamma$}
\\ 
\cline{2-3}
\cline{4-5}
\cline{6-7}
\ifproofpre{\\[-8pt]}{\\[-24pt]} % for space below clines
&
\multicolumn{1}{c}{$\tme{f}{\calM'}{i}$} & \multicolumn{1}{c}{$a$} &
\multicolumn{1}{c}{$\tme{f}{\calM'}{i}$} & \multicolumn{1}{c}{$a$} &
\multicolumn{1}{c}{$\tme{f}{\calM'}{i}$} & \multicolumn{1}{c}{$a$} 
&
\multicolumn{1}{c}{\smash{\raisebox{10pt}{$\dfrac{\gamma\gamma_4\rightarrow\gamma}{\gamma\rightarrow g}$}}}
&
\multicolumn{1}{c}{\smash{\raisebox{10pt}{$\dfrac{\gamma\gamma_0\rightarrow\gamma}{\gamma\rightarrow g}$}}}
\\[2pt] % for space above hline
\hline
$\chi=50$$^a$
& 0.42 & 0.025 & \sim0.6^a & \sim0.03 & \sim0.5^a& \sim0.03 & \sim1.4^a& \sim1.1^a\\
$\chi=100$
& 0.30 & 0.012 & 0.43 & 0.012 & 0.30 & 0.018 & 1.44 & 1.01\\
$\chi=200$
& 0.23 & 0.007 & 0.33 & 0.007 & 0.23 & 0.009 & 1.43 & 1.00\\
Harmonic & 
\multicolumn{1}{c}{---} & & \multicolumn{1}{c}{---} & & \multicolumn{1}{c}{---} & &
1.41 & 1
\end{tabular}
\end{ruledtabular}
\raggedright
$^a$ The $\gamma\gamma\rightarrow\gamma$
intrinsic matrix elements for $\chi=50$ can only
be crudely approximated, since the Mikhailov plot yields values which are
not strongly linear [Fig.~\ref{fig-gammarotor-axial-mikhailov}(d,g)].
The estimated 
parameters used in the analysis are 
$M_1\approx0.9$ for $\gamma\gamma_4\rightarrow\gamma$ and  
$M_1\approx0.4$ for
$\gamma\gamma_0\rightarrow\gamma$.
\end{center}
\end{table*}
%----------------------------------------------------------------

%----------------------------------------------------------------

A more comprehensive and meaningful examination of electromagnetic
transition strengths is realized by considering the interband
transitions in aggregate, according to the Mikhailov mixing
formalism~\cite{mikhailov1966:mixing-APS}.  Within this framework, all
transition amplitudes are expressed in terms of a single intrinsic
electromagnetic matrix element and single mixing parameter between
each pair of bands.  The amplitudes are expected to fall on a straight line
on an appropriate (Mikhailov) plot of $\trme{K_2J_2}{\calM}{K_1J_1}$
or, commonly, $B(E2)^{1/2}$ \textit{vs.}  $J_2(J_2+1)-J_1(J_1+1)$.
The intrinsic matrix elements and mixing parameter are identified from
the slope and intercept.

Specifically, for interband transitions  with $\Delta K=2$,
 the leading-order band
mixing relation for $E2$ reduced matrix
elements is~\cite[(4-210)]{bohr1998:v2}
\begin{multline}
\label{eqn-mikhailov-DeltaK2}
\trme{K_2J_2}{\calM}{K_1J_1} = \sigma_1 (2J_1+1)^{1/2}
\tcg{J_1}{K_1}{2}{2}{J_2}{K_2}
\ifproofpre{\\\times}{}
\bigl[M_1+M_2[J_2(J_2+1)-J_1(J_1+1)]\bigr],
\end{multline}
where it is assumed that $K_2=K_1+2$, and where
$\sigma_1=\sqrt{2}$ if $K_1=0$ or $\sigma_1=1$ otherwise.  The
parameters in this expression are related to the intrinsic matrix
element $\tme{K_2}{\calM'}{K_1}$, mixing matrix element
$\tme{K_2}{\varepsilon_{+2}}{K_1}$, and intrinsic quadrupole moment
$Q_0$ by $M_1=\tme{K_2}{\calM'}{K_1}-4(K_1+1)M_2$ and
$M_2=[15/(8\pi)]^{1/2} e
Q_0\tme{K_2}{\varepsilon_{+2}}{K_1}$~\cite[(4-211)]{bohr1998:v2}.  The
intrinsic matrix element may thus be extracted from the slope and
intercept as
\begin{equation}
\label{eqn-mikhailov-DeltaK2-IME}
\tme{K_2}{\calM'}{K_1}=M_1+4(K_1+1)M_2.
\end{equation}
More specific expressions for $K$-\textit{decreasing} and
$K$-\textit{increasing} transitions, in terms of $B(E2)$ reduced
transition probabilities, are given in Appendix~\ref{app-mikhailov}.

The interband quadrupole transition strengths for the Bohr Hamiltonian
calculations of Sec.~\ref{sec-axial-spectra} are shown in Fig.~\ref{fig-gammarotor-axial-mikhailov} in
Mikhailov form.  They are plotted as $B(E2)^{1/2}$ \textit{vs.}
$L_f(L_f+1)-L_i(L_i+1)$, for transitions between states with $L\leq
6$.  For the most part, the transition amplitudes do indeed follow an
essentially linear pattern, and it is therefore meaningful to extract
effective intrinsic matrix elements, well as mixing parameters, from
the Mikhailov analysis.  (The Mikhailov formalism has been applied to
extract effective intrinsic matrix elements from the interacting boson
model~\cite{iachello1987:ibm}, in a similar fashion, in
Refs.~\cite{warner1981:168er-ibm,aprahamian2006:162dy-grid}.)
However, deviations from a linear relation are significant for
transitions involving the two-phonon quasi-bands for $\chi=50$
[Fig.~\ref{fig-gammarotor-axial-mikhailov}(left)], as might be
expected from the substantial $\gamma$-softness and deviations from
rotational \textit{energy} spacings already noted for these bands.
The resulting intrinsic matrix elements for the $\gamma\rightarrow g$,
$\gamma\gamma_4\rightarrow\gamma$, and
$\gamma\gamma_0\rightarrow\gamma$ transitions, obtained
from~(\ref{eqn-mikhailov-decr-IME})
and~(\ref{eqn-mikhailov-incr-IME}), are listed in
Table~\ref{tab-ime-axial}, together with the dimensionless mixing
parameter $a=\abs{M_2/M_1}$ (see Appendix~\ref{app-mikhailov}).  The
normalization of the electric quadrupole operator $\calM(E2)$ is
arbitrary in the present analysis.  To provide a scale for comparison
with experiment, the intrinsic matrix elements in
Table~\ref{tab-ime-axial} are given relative to the square root of the
in-band $B(E2;2^+_1\rightarrow0^+_1)$.

For harmonic $\gamma$ vibration, the ratios of the
$\gamma\gamma\rightarrow\gamma$ intrinsic matrix elements to the
$\gamma\rightarrow g$ intrinsic matrix element are expected to be
$\tme{\gamma}{\calM'}{\gamma\gamma_4}/\tme{g}{\calM'}{\gamma}=\sqrt{2}\approx1.41$
and $\tme{\gamma}{\calM'}{\gamma\gamma_0}/\tme{g}{\calM'}{\gamma}=1$,
according to the proportion noted in Sec.~\ref{sec-axial-evoln}.  For
comparison, ratios of the intrinsic matrix elements extracted from the
Bohr Hamiltonian numerical calculations are given in the last two
columns of Table~\ref{tab-ime-axial}.  Note the rapid quantitative
approach of these calculated ratios to the expected harmonic values.
Even the $\gamma\gamma\rightarrow\gamma$ transitions for the soft
$\chi=50$ case are essentially consistent with harmonic ratios, to the
extent that slope and intercept parameters can meaningfully be
extracted in this instance
[Fig.~\ref{fig-gammarotor-axial-mikhailov}(d,g)].  For $\chi=200$,
harmonic values are obtained to within $\sim1\%$.

The bandmixing, indicated by the Mikhailov plot
\textit{slopes}, is substantial in all the
cases considered in Table~\ref{tab-ime-axial}.  The harmonicity of
the intrinsic matrix elements is therefore \textit{not} apparent
simply from the plot \textit{intercepts} buy only 
after the leading-order bandmixing
corrections~(\ref{eqn-mikhailov-decr-IME})
and~(\ref{eqn-mikhailov-incr-IME}) are taken into account.  For
example, even for the most adiabatic case, $\chi=200$, the
$\gamma\rightarrow g$
[Fig.~\ref{fig-gammarotor-axial-mikhailov}(c)]and
$\gamma\gamma_4\rightarrow\gamma$
[Fig.~\ref{fig-gammarotor-axial-mikhailov}(f)] Mikhailov plots both
have slope parameters $a\approx0.012$, resulting in a $5\%$ adjustment
to the $\gamma\rightarrow g$ intrinsic matrix element and a $14\%$
adjustment to the $\gamma\gamma_4\rightarrow\gamma$ intrinsic matrix
element.  

In summary, although the strengths of the individual interband
\textit{transitions} only approach the limit of an adiabatic rotor (and,
more specificially, harmonic vibration) \textit{gradually}, as
observed from Fig.~\ref{fig-gammarotor-evoln}(d), this deviation is
quantitatively well-described in terms of a \textit{rapid} approach to
harmonic values of the interband \textit{intrinsic} matrix elements,
but with the individual transition strengths modified by leading-order
$\Delta K=2$ bandmixing~(\ref{eqn-mikhailov-DeltaK2}).  The strength
of this mixing then gradually decreases with increasing $\gamma$ stiffness.

\section{Onset of rigid triaxiality}
\label{sec-triax}
%----------------------------------------------------------------
\begin{figure*}
\begin{center}
\includegraphics*[width=0.95\hsize]{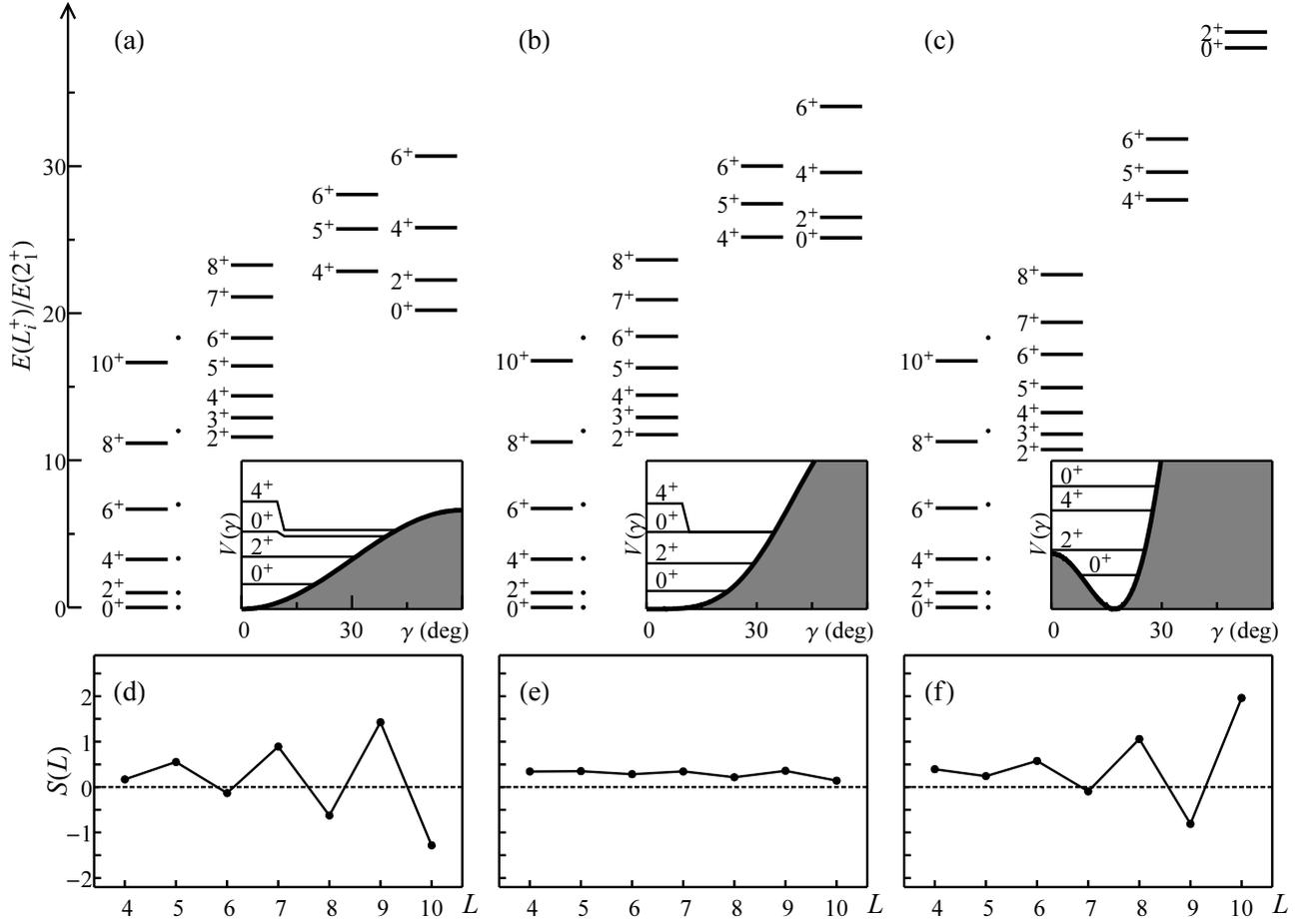}
\end{center}
\caption{Level schemes for the angular
Hamiltonian~(\ref{eqn-gammarotor-xi}), for 
(a)~$\xi=0$ with $\chi=50$, (d)~$\xi=0.5$ with $\chi=100$, and
(e)~$\xi=0.8$ with $\chi=500$.  Rotational
$L(L+1)$ energies for the yrast band are indicated by the dots. The potential
$V(\gamma)$ is shown in the inset, with the ground, quasi-$\gamma$, and quasi-$\gamma\gamma$ bandhead 
energies indicated.  (d--f)~Staggering of level energies within the
quasi-$\gamma$ band, as measured by
the energy second difference $S(L)$.  Figure adapted from Ref.~\cite{caprio2009:gammaharmonic}.
}
\label{fig-gammarotor-triax-schemes}
\end{figure*}
%----------------------------------------------------------------
%----------------------------------------------------------------
\begin{figure*}
\begin{center}
\includegraphics*[width=0.7\hsize]{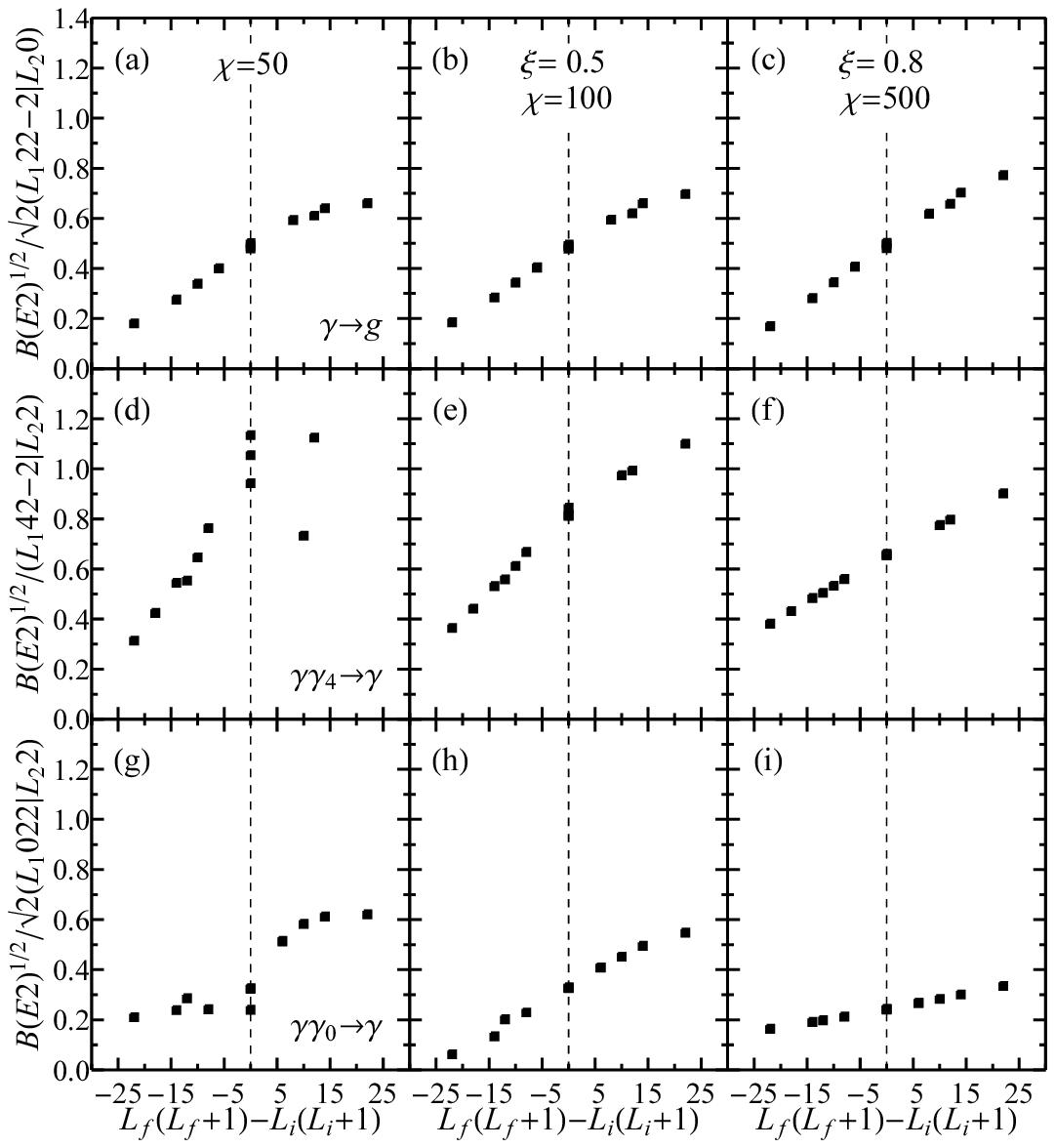}
\end{center}
\caption{
Interband transition amplitudes $B(E2)^{1/2}$, from the $\gamma$ quasi-band to the
ground state band~(top), from the $\gamma\gamma_4$ quasi-band
to the $\gamma$ quasi-band~(middle), and from the $\gamma\gamma_0$ quasi-band
to the $\gamma$ quasi-band~(bottom), for Mikhailov analysis.  Plots
are included for the calculations of Fig.~\ref{fig-gammarotor-triax-schemes}, with
$\xi=0$ ($\chi=50$)~(left), $\xi=0.5$ ($\chi=100$)~(middle),
and $\xi=0.8$ ($\chi=500$)~(right).  The values shown are for
transitions between levels with $L\leq6$, normalized to
$B(E2;2^+_1\rightarrow0^+_1)\equiv1$.  Figure panels~(a--f) adapted from Ref.~\cite{caprio2009:gammaharmonic}.
}
\label{fig-gammarotor-triax-mikhailov}
\end{figure*}
%----------------------------------------------------------------
%----------------------------------------------------------------
%----------------------------------------------------------------
\begin{table*}
  \caption{Electric quadrupole interband intrinsic matrix elements $\tme{f}{\calM'}{i}$ 
    and mixing parameters, for different
    $\gamma$ potential shapes chosen to reproduce the onset of weak rigid triaxiality, as extracted from
    the Mikhailov analyses of Fig.~\ref{fig-gammarotor-triax-mikhailov}.  Ratios, as indicators of 
    anharmonicity, are tabulated in the final two
    columns.  The $\grpy{5}$
    triaxial estimate~\cite{iachello2003:y5} is included for comparison.  The values
    for the intrinsic matrix elements are normalized to 
    $B(E2;2^+_1\rightarrow0^+_1)\equiv1$.
    The results in this table also serve to correct intrinsic
    matrix element values given previously in
    Table~1 of Ref.~\cite{caprio2009:gammatriax}.  The roles of initial
and final bands were interchanged, in that analysis, when extracting the slope parameter
from~(\ref{eqn-mikhailov-decr}) and~(\ref{eqn-mikhailov-incr}), 
resulting in the use of an incorrect sign for the bandmixing correction term
in~(\ref{eqn-mikhailov-decr-IME}) and~(\ref{eqn-mikhailov-incr-IME}).
}
\label{tab-ime-triax}
\begin{ruledtabular}
\begin{tabular}{ldddddddd}
\ifproofpre{\\[-8pt]}{\\[-24pt]} % for fractions
 &
\multicolumn{2}{c}{$\gamma\rightarrow g$}
&
\multicolumn{2}{c}{$\gamma\gamma_4\rightarrow\gamma$}
&
\multicolumn{2}{c}{$\gamma\gamma_0\rightarrow\gamma$}
\\ 
\cline{2-3}
\cline{4-5}
\cline{6-7}
\ifproofpre{\\[-8pt]}{\\[-24pt]} % for space below clines
&
\multicolumn{1}{c}{$\tme{f}{\calM'}{i}$} & \multicolumn{1}{c}{$a$} &
\multicolumn{1}{c}{$\tme{f}{\calM'}{i}$} & \multicolumn{1}{c}{$a$} &
\multicolumn{1}{c}{$\tme{f}{\calM'}{i}$} & \multicolumn{1}{c}{$a$} 
&
\multicolumn{1}{c}{\smash{\raisebox{8pt}{$\dfrac{\gamma\gamma_4\rightarrow\gamma}{\gamma\rightarrow g}$}}}
&
\multicolumn{1}{c}{\smash{\raisebox{8pt}{$\dfrac{\gamma\gamma_0\rightarrow\gamma}{\gamma\rightarrow g}$}}}
\\[2pt] % for space above hline
\hline
$\xi=0.5$ ($\chi=100$)
& 0.43 & 0.025 & 0.58 & 0.022 & 0.37 & 0.035 & 1.36 & 0.87\\
$\xi=0.8$ ($\chi=500$)
& 0.43 & 0.028 & 0.51 & 0.018 & 0.27 & 0.015 & 1.18 & 0.63\\
$\grpy{5}$ & 
\multicolumn{1}{c}{---} & & \multicolumn{1}{c}{---} & & \multicolumn{1}{c}{---} & &
1.23 & 0.73
\end{tabular}
\end{ruledtabular}
\end{table*}
%----------------------------------------------------------------

%----------------------------------------------------------------

The excitation spectrum may be expected to change dramatically with
the onset of rigid triaxiality.  The Bohr Hamiltonian predictions
ultimately approach a $\gamma=30^\circ$ Davydov rotor
spectrum~\cite{davydov1958:arm-intro} for confinement by a
sufficiently stiff $\cos^2 3\gamma$ potential~\cite{rowe2009:acm}.
However, the initial onset of triaxiality is reflected in much more
subtle deviations from the characteristics of an axially symmetric
rotor.  The difference between axial and triaxial minima in the
potential is obscured by the substantial dynamical fluctuations in
$\gamma$ present in both cases.  As noted in Sec.~\ref{sec-method},
the onset of triaxiality may be investigated by considering the
introduction of a $\cos^2 3\gamma$ contribution,
\textit{i.e.}, nonzero $\xi$, in the
Hamiltonian~(\ref{eqn-gammarotor-xi}).

The results of calculations for two representative potentials are
shown in Fig.~\ref{fig-gammarotor-triax-schemes}: the soft or
``critical'' axial minimum ($\xi=0.5$)
[Fig.~\ref{fig-gammarotor-triax-schemes}(b)] and a weakly triaxial
minimum ($\xi=0.8$) [Fig.~\ref{fig-gammarotor-triax-schemes}(c)].  For
each of these calculations, the potential depth, or $\chi$, is chosen
to give $E(2^+_\gamma)/E(2^+_1)\approx10$, again appropriate to the
well-deformed rare earth nuclei.  The comparable axial rotor
calculation with the same $\gamma$ band energy, \textit{i.e.},
$\chi=50$, is shown again as a baseline for comparison
[Fig.~\ref{fig-gammarotor-triax-schemes}(a)].

In Fig.~\ref{fig-gammarotor-triax-schemes}, the
$\gamma$-phonon quasiband structure is seen to remain intact.  Our concern
is therefore with the principal
spectroscopic properties of these bands~--- excitation energies of the
bands, deviations from rotational energy spacing within the bands, and
electric quadrupole intrinsic matrix elements.  The two-phonon energy
anharmonicities evolve from slightly negative
($E_{\gamma\gamma}/E_{\gamma}<2$) for $\xi=0$
[Fig.~\ref{fig-gammarotor-triax-schemes}(a)] to positive
($E_{\gamma\gamma}/E_{\gamma}>2$)
[Fig.~\ref{fig-gammarotor-triax-schemes}(b,c)] with the introduction
of triaxial tendencies.  The anharmonicity of the
$\gamma\gamma_0$ band rises more rapidly than that of the $\gamma\gamma_4$
quasi-band.  Qualitatively, this is consistent with evolution towards
a $\gamma$-stiff, adiabatic triaxial
rotor~\cite{rowe2010:rowanwood}, for which the $K=4$
quasi-band is a triaxial \textit{rotational} excitation and the $K=0$
quasi-band is a $\gamma$ \textit{vibrational} excitation.

The level energies within the $\gamma$ band progress, with increasing
$\xi$, from $\gamma$-soft staggering [$2(34)(56)\ldots$] to the
reverse pattern associated with triaxial rotation
[$(23)(45)\ldots$]~\cite{davydov1958:arm-intro}.  As in
Sec.~\ref{sec-axial-spectra}, the staggering may be seen most
immediately from plots of the second difference $S(L)$
[Fig.~\ref{fig-gammarotor-triax-schemes}(d--f)], which has minima at
even $L$ for $\gamma$-soft staggering or at odd $L$ for triaxial
staggering.  

The ``centrifugal stretching'' phenomenon in the yrast band,
\textit{i.e.}, reduction of $E(L^+_1)/E(2^+_1)$ relative to $L(L+1)$
spacing, persists [Fig.~\ref{fig-gammarotor-triax-schemes}(b,c)] at
about the same level as for $\chi=50$.  However, the growth in
rotational constant (and general deviation from rotational behavior)
for the excited, especially $\gamma\gamma$, bands is tamed relative to
the axial calculation.  This may be at least qualitatively understood
by comparing the potential plots in
Fig.~\ref{fig-gammarotor-triax-schemes}(a--c,insets).  The axial
calculation of Fig.~\ref{fig-gammarotor-triax-schemes}(a),
as noted in Sec.~\ref{sec-axial-spectra}, provides only weak
confinement at the $\gamma\gamma$ band energies
($\gamma\lesssim40^\circ$).  Although the nominally ``softer'' calculation of
Fig.~\ref{fig-gammarotor-triax-schemes}(b) does provide weaker
confinement, compared to this axial calculation, at the \textit{ground
  state} energy, it actually provides stiffer confinement, to a
smaller range of $\gamma$ values ($\gamma\lesssim30^\circ$), at the
$\gamma\gamma$ band energies.  [This effect may be more properly considered a reflection
of the steep rise in the $\cos^2 3\gamma$ term used to create the
triaxial confinement than an intrinsic property of the onset of
triaxiality \textit{per se}. There is no inherent calculational reason
not to consider a potential with, for instance, a triaxial minimum
located at the same position as in Fig.~\ref{fig-gammarotor-triax-schemes}(c,inset) but a
lower barrier at $\gamma=60^\circ$.\fngenpotl]  A similar
observation may be made for the calculation of
Fig.~\ref{fig-gammarotor-triax-schemes}(c), which provides confinement
to triaxial $\gamma$ at the ground state energy, but simply provides
(axial) confinement to $\gamma\lesssim30^\circ$ at the $\gamma\gamma$
band energies.

For the weakly triaxial calculations considered here, the interband
transition strengths continue to follow an essentially linear pattern
on a Mikhailov plot, as expected for rotational bandmixing, as seen in
Fig.~\ref{fig-gammarotor-triax-mikhailov}.  The
$\gamma\gamma\rightarrow\gamma$ transitions, in fact, demonstrate
better linear  behavior
 [Fig.~\ref{fig-gammarotor-triax-mikhailov}(e--f,h--i)]
than for $\chi=50$ [Fig.~\ref{fig-gammarotor-triax-mikhailov}(d,g)].
Interband intrinsic matrix elements may therefore again be extracted
from the Mikhailov analysis, as given in
Table~\ref{tab-ime-triax}.  The $\gamma\rightarrow g$ intrinsic matrix
element remains essentially constant, and equal to that for the axial
$\chi=50$ calculation, but the $\gamma\gamma_4\rightarrow \gamma$, and
$\gamma\gamma_0\rightarrow\gamma$ intrinsic matrix elements decrease
substantially compared to the harmonic $\gamma$-vibrational values.

Such a reduction of the $\gamma\gamma\rightarrow\gamma$ intrinsic
matrix elements, relative to the harmonic values, has already been
proposed~\cite{iachello2003:y5} on
relatively simple grounds.  Supposing an adiabatic separation of
rotation from vibration, and furthermore imposing a small-$\gamma$ approximation,
yields a one-dimensional Schr\"odinger equation problem in $\gamma$.
In Ref.~\cite{iachello2003:y5}, a square well
is then adopted for $V(\gamma)$ to simulate the onset of triaxiality.  This
yields the $\grpy{5}$ estimate shown for comparison in
Table~\ref{tab-ime-triax}.

\section{Wave function probability distributions}
\label{sec-decomp}
%----------------------------------------------------------------
\begin{figure*}
\begin{center}
\includegraphics*[width=0.95\hsize]{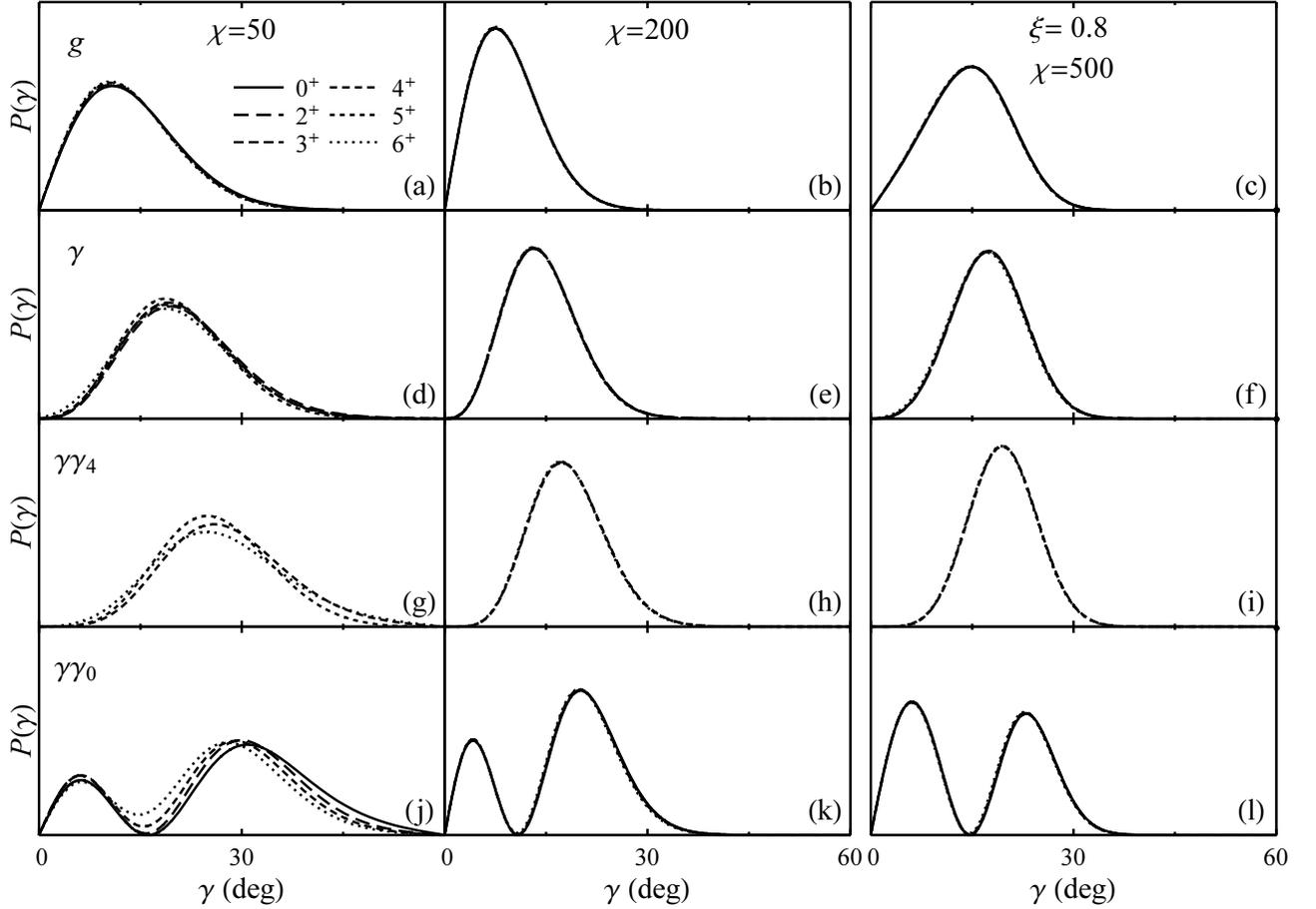}
\end{center}
\caption{
Probability distributions with respect to $\gamma$ for low-lying quasi-band members in calculations with
Hamiltonian~(\ref{eqn-gammarotor-xi}), for the axial cases
$\chi=50$ (left) and $\chi=200$
(middle), both with $\xi=0$, and for the weakly triaxial case $\xi=0.8$ with $\chi=500$ (right).  Probability
distributions are shown for members of the ground
state, $\gamma$, $\gamma\gamma_4$, and $\gamma\gamma_0$ quasi-bands (top to
bottom, respectively), with $L\leq 6$. 
}
\label{fig-gammarotor-combo-Pgamma}
\end{figure*}
%----------------------------------------------------------------
%----------------------------------------------------------------
\begin{figure*}
\begin{center}
\includegraphics*[width=0.95\hsize]{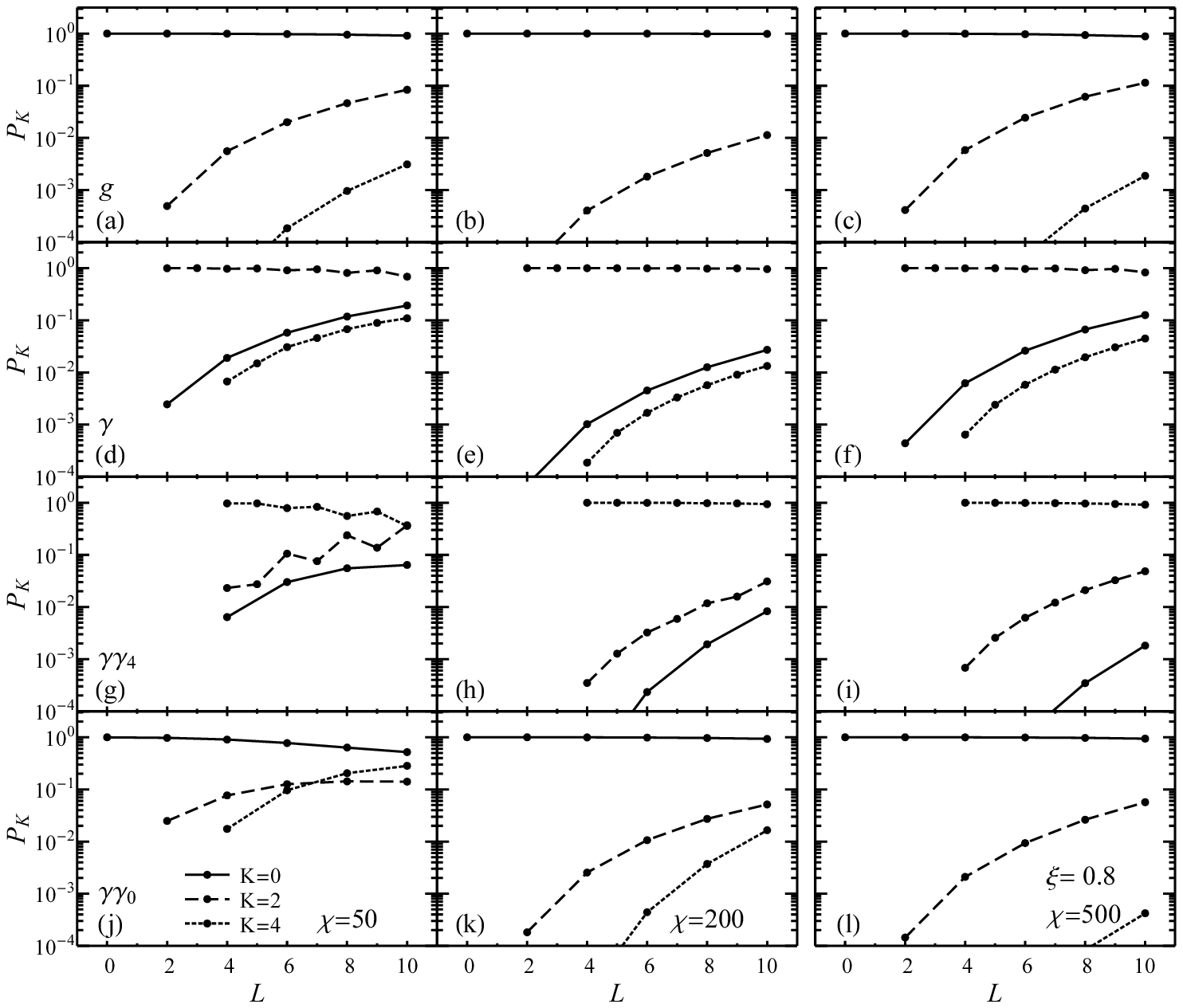}
\end{center}
\caption{
The $K$ content of low-lying quasi-band members in calculations with
Hamiltonian~(\ref{eqn-gammarotor-xi}), for the axial cases
$\chi=50$ (left) and $\chi=200$
(middle), both with $\xi=0$, and for the weakly triaxial case $\xi=0.8$ with $\chi=500$ (right).  Probabilities $P_K$ for
$K=0$ (solid curve), $K=2$ (dashed curve), and $K=4$ (dotted curve) are shown for members of the ground
state, $\gamma$, $\gamma\gamma_4$, and $\gamma\gamma_0$ quasi-bands (top to
bottom, respectively), with $L\leq 10$. 
}
\label{fig-gammarotor-combo-PK}
\end{figure*}
%----------------------------------------------------------------

In the limit of adiabatic separation of the $\gamma$ and rotational
degrees of freedom, the wave functions of all members of a band would
be given by
\begin{equation} 
\label{eqn-intrinsic}
\psi_{KLM}(\gamma,\Omega)=
F_{K}(\gamma) \xi^{(L)}_{KM}(\Omega),
\end{equation} 
where the function $F_K(\gamma)$ would be identical for all states
within the same band, independent of $L$.  The band is characterized
by intrinsic angular momentum projection $K$.  This may be contrasted
to the general situation~(\ref{eqn-expansion-F}), in which all even
$K$ with $0\leq K \leq L$ (or $2\leq K\leq L$ for $L$ odd) can
contribute, and the coefficients $F_K(\gamma)$ need not be directly
related for different states.  The breaking of adiabaticity
has already been seen to have spectroscopic consequences
(Secs.~\ref{sec-axial} and~\ref{sec-triax}).  Here we shall more
directly inspect the wave functions themselves, through the
probability distributions.  Specifically, we examine the probability
distribution $P(\gamma)$, with respect the $\gamma$ coordinate, after
integration over Euler angles, and the probability decomposition
$P_K$, with respect to the $K$ quantum number for the Euler angle
(rotational) dependence, after integration over $\gamma$.  The
calculational details are given in Appendix~\ref{app-prob}.

First, considering $P(\gamma)$, results are given in
Fig.~\ref{fig-gammarotor-combo-Pgamma} for 
the softest axial
calculation of Sec.~\ref{sec-axial} ($\chi=50$)
[Fig.~\ref{fig-gammarotor-combo-Pgamma}(left)], the stiffest axial
calculation of Sec.~\ref{sec-axial} ($\chi=200$)
[Fig.~\ref{fig-gammarotor-combo-Pgamma}(middle)], and the weakly
triaxial calculation of Sec.~\ref{sec-triax} ($\xi=0.8$ with $\chi=500$)
[Fig.~\ref{fig-gammarotor-combo-Pgamma}(right)].  Successive panels
(top to bottom) show the $P(\gamma)$ distributions for the ground,
$\gamma$, $\gamma\gamma_4$, and $\gamma\gamma_0$ band members,
respectively, with $L\leq6$.  All the $P(\gamma)$ vanish at
$\gamma=0^\circ$ and $\gamma=60^\circ$, due to the volume element for
the Bohr coordinates (see Appendix~\ref{app-prob}).

The basic features seen in Fig.~\ref{fig-gammarotor-combo-Pgamma} may
be qualitatively understood in terms of the small-$\gamma$ limit
of~(\ref{eqn-gammarotor-xi}), which reduces (\textit{e.g.},
Ref.~\cite{iachello2003:y5}) to a two-dimensional harmonic oscillator
problem, with two-dimensional angular momentum $m=K/2$ and with
$\gamma$ as the ``radial'' variable.  The $K=2n_\gamma$ (or
$m=n_\gamma$) bands, \textit{i.e.}, the ground, $\gamma$, and
$\gamma\gamma_4$ bands, have probability distributions which are
nodeless.  These move towards higher $\gamma$ with increasing phonon
number $n_\gamma$ [Fig.~\ref{fig-gammarotor-combo-Pgamma}(a,d,g) or
Fig.~\ref{fig-gammarotor-combo-Pgamma}(b,e,h)].  The centers of the
probability distributions are at substantially nonzero $\gamma$
values, in the $10^\circ$--$30^\circ$ range,
but move towards smaller $\gamma$ for larger stiffess [compare
Fig.~\ref{fig-gammarotor-combo-Pgamma}(left) with
Fig.~\ref{fig-gammarotor-combo-Pgamma}(middle)].  All these properties
are as anticipated from the $\gammabar$ values in
Fig.~\ref{fig-gammarotor-axial-gamma}.  For the $\gamma\gamma_0$ band,
which is characterized by $K=2(n_\gamma-2)$ (or
$m=n_\gamma-2$), the probability distributions have a single node 
[Fig.~\ref{fig-gammarotor-combo-Pgamma}(j,k)].

Adiabatic separation~(\ref{eqn-intrinsic}) implies identical
$P(\gamma)$ distributions for all members of the same band.  Indeed,
the $P(\gamma)$ curves are virtually indistinguishable between band
members for the examples in Fig.~\ref{fig-gammarotor-combo-Pgamma}.
The exceptions are, once again, the $\gamma\gamma$ bands in the
$\chi=50$ calculation [Fig.~\ref{fig-gammarotor-combo-Pgamma}(g,j)].
There is some slight displacement between the curves for the different
members of the ground or $\gamma$ bands in this calculation as well.
The breaking of adiabaticity is also apparent for the $\gamma\gamma_0$
band members with $L>0$, from the disappearance of the node in
$P(\gamma)$, which indicates that multiple $K$ values must contribute
to the wave function.\fnKnode

It is interesting to note the qualitative differences of the more
triaxial calculation [Fig.~\ref{fig-gammarotor-combo-Pgamma}(right)]
from the axial calculations
[Fig.~\ref{fig-gammarotor-combo-Pgamma}(left,middle)].
The $P(\gamma)$ for the ground, $\gamma$, and $\gamma\gamma_4$ bands (\textit{i.e.}, those
with nodeless distributions)
[Fig.~\ref{fig-gammarotor-combo-Pgamma}(c,f,i)] are peaked at $\gamma$
values roughly comparable to those for the $\chi=50$ ``axial'' calculation
[Fig.~\ref{fig-gammarotor-combo-Pgamma}(a,d,g)] (recall that the
parameters were chosen so that these calculations share the same
$\gamma$ band energy) but are more sharply peaked.  The 
$\gamma\gamma_0$ distribution
[Fig.~\ref{fig-gammarotor-combo-Pgamma}(l)] shows a marked enhacement
of the peak at \textit{small} (axial) $\gamma$.  This may seem
counterintuitive for a ``triaxial'' calculation, but, as already
remarked in Sec.~\ref{sec-triax}, the triaxial confinement is limited
to the ground state band energy.\fntriaxortho

In interpreting the $P(\gamma)$ distributions as indicators of
adiabaticity, it should be noted that, although adiabatic separation
implies identical $P(\gamma)$ distributions, the converse is
\textit{not} strictly true.  Adiabaticity might be violated, and
several $K$ values might contribute in~(\ref{eqn-expansion-F}), but
the various $F_K(\gamma)$ for the different band members
may be related such that, nonetheless, the same $P(\gamma)$
distributions are obtained after integration over Euler angles.
Therefore, these distributions can only be conclusively taken to
indicate adiabaticity if it is also known that only one $K$ value
contributes significantly.

The contributions of different $K$ values in each of the bands
(ground, $\gamma$, $\gamma\gamma_4$, and $\gamma\gamma_0$) are shown
in Fig.~\ref{fig-gammarotor-combo-PK}, for each band member with
$L\leq10$. For the calculations in Fig.~\ref{fig-gammarotor-combo-PK},
the bandhead states have essentially pure $K$.  The largest admixture
in a bandhead state is $\sim3\%$ for the $\gamma\gamma_4$ bandhead in
the $\chi=50$ calculation, but the bandhead $K$ admixtures in the
other calculations are all $<10^{-3}$.  (Note that the
$\gamma\gamma_0$ bandhead, as an $L=0$ state, trivially has pure
$K=0$.)  The admixtures increase with $L$ within each band.  Again,
the extremes are in the $\gamma\gamma$ bands for $\chi=50$, where the
admixtures account for approximately half the probability at $L=10$
[Fig.~\ref{fig-gammarotor-combo-PK}(g,j)].  In contrast, for the
weakly triaxial calculation
[Fig.~\ref{fig-gammarotor-combo-PK}(right)], the $K$ admixtures in the
$\gamma\gamma$ bands are actually slightly \textit{smaller} than for
the ground state band.  Indeed, they closely match the $K$ admixtures
of the corresponding bands in the stiff axial $\chi=200$ calculation
[Fig.~\ref{fig-gammarotor-combo-PK}(middle)].  This observation is
consistent with the characterization of these bands as relatively
``good'' axial rotational bands, as suggested spectroscopically in
Sec.~\ref{sec-triax}.

\section{Conclusion}
\label{sec-concl}

The possibility of exact diagonalization of the Bohr Hamiltonian for
essentially arbitrary $\beta$ and $\gamma$ stiffness opens the door
for direct comparison of the Bohr Hamiltonian predictions with
experiment throughout the range of possible dynamics for the nuclear
quadrupole degree of freedom.  At a phenomenological level, this
permits meaningful tests of the Bohr Hamiltonian for general
rotor-vibrator nuclei.

For instance, in the past, interpretation of rotational ``phonon''
states, although nominally within the Bohr description, has largely
been at a schematic level (\textit{e.g.},
Refs.~\cite{gunther1967:166er-168er-beta,riedinger1969:152sm154gd-beta,warner1981:168er-ibm,wu1996:gamma-fragmentation-os,haertlein1998:168er-coulex}):
adiabatic separation of the rotational and vibrational degrees of
freedom is assumed, the $\beta$ and $\gamma$ excitations are taken to
be harmonic, and phonon selection rules are assumed for electric
quadrupole transitions.  These predictions are then adjusted by the
leading-order spin-dependent bandmixing relation, but with \textit{ad
hoc} mixing parameters.
Here, instead, we explore exact predictions of the Bohr Hamiltonian,
both for axial and weakly triaxial confinement.  

The present analysis, which has been restricted to the $\gamma$ and
rotational degrees of freedom, provides a starting point for understanding
the full dynamics involving all five Bohr coordinate degrees of
freedom, \textit{i.e.}, considering coupling with the $\beta$ degree
of freedom as well.  Many of the qualitative properties of the present
solution may be expected to carry over (see,
\textit{e.g.}, Fig.~4 of Ref.~\cite{caprio2009:gammatriax}).  However, the
introduction of $\beta$ softness may generally be expected to
quantitatively alter the results, for instance, further attenuating
the rotational character of the bands [\textit{e.g.}, reducing the
ratio $E(4^+_1)/E(2^+_1)$]. Moreover, in the case of near degeneracy
of the $\gamma$ phonon or multiphonon bands with bands involving
$\beta$ excitations, bandmixing can substantially alter the results.
Therefore, detailed comparison with experiment should be made in the
context of a full treatment incorporating $\beta$ softness.

Microscopic descriptions of nuclear collectivity rely upon a reduction
of the many-body problem to one involving effective collective degrees
of freedom.  Mean-field approaches to deriving the quadrupole
collective dynamics (reviewed in, \textit{e.g.},
Refs.~\cite{klein1991:collective,prochniak2009:bohr-collective-hf-hg,matsuyanagi2010:micro-collective-problems})
yield a Hamiltonian involving a much more general,
coordinate-dependent form for the kinetic energy operator than the
conventional but schematic Laplacian form considered
in~(\ref{eqn-Hbohr}).  The resulting \textit{generalized Bohr
Hamiltonian}~\cite{prochniak2009:bohr-collective-hf-hg} may be
represented in terms of coordinate-dependent moments of inertia.  It
should be noted that the ACM can readily accommodate Hamiltonians
involving much more general differential
operators~\cite{rowe2005:radial-me-su11} in the $\beta$ and angular
variables than the simple Laplacian form. For instance, scalar-coupled
products of the quadrupole momentum tensor $p$ and coordinate tensor
$q$ constitute an important special case considered in the geometric
collective model~\cite{hess1980:gcm-details-238u,eisenberg1987:v1}.
The Bohr kinetic energy is obtained as the lowest-order term $(p\times
p)^{(0)}$, and attention in phenomenological studies has largely been
limited to the next term $(p\times q \times p)^{(0)}$.  These and
higher-order terms in the coordinate dependence may be combined to
recover much or all of the flexibility of the generalized Bohr
Hamiltonian~\cite{jolos-UNP}.

Even further generalizations may be required.  For instance, the
$\grpsp{3,\bbR}$ symplectic shell model framework gives rise to a
collective model in which the generalized Bohr Hamiltonian must be
augmented with a vorticity degree of
freedom~\cite{rowe1985:micro-collective-sp6r}.  Since the collective
model serves as the intermediate link between microscopic theories
and spectroscopic predictions, it is essential to determine the
limitations of the Bohr Hamiltonian and the nature of the
modifications required such that its predictions can accurately
describe the observed phenomena.

\begin{acknowledgments}
Valuable discussions with N.~V.~Zamfir, 
D.~J.~Rowe, S.~De Baerdemacker, F.~Iachello, S.~Frauendorf, and A.~Aprahamian are
gratefully acknowledged.  This work was supported by the US DOE under
grant DE-FG02-95ER-40934.
\end{acknowledgments}

\appendix

\section{Restriction to angular coordinates}
\label{app-angular}

In this appendix, the reduction of the full Bohr
Hamiltonian~(\ref{eqn-Hbohr}) to an angular
Hamiltonian~(\ref{eqn-gammarotor}) is briefly summarized.  First, for
convenience, let us simplify the Bohr Hamiltonian to its equivalent
dimensionless form
\begin{equation}
\label{eqn-Hbohr-dimless}
H=-\biggl[\Deltahat-\frac{\Lambdahat^2}{\beta^2}\biggr]+V(\beta,\gamma),
\end{equation}
where
\begin{equation}
\label{eqn-Deltahat}
\Deltahat=\frac{1}{\beta^4}\frac{\partial}{\partial\beta}\beta^4\frac{\partial}{\partial\beta},
\end{equation}
by rescaling $H\rightarrow (2B/\hbar^2)H$ and  $V\rightarrow (2B/\hbar^2)V$.
The two routes to obtaining an angular Hamiltonian indicated in
Sec.~\ref{sec-method-hamiltonian} proceed more precisely as follows:

(1)~Schematically, rigid $\beta$ deformation ($\beta\approx\beta_0$)
is obtained if the nuclear wave function $\Psi(\beta,\gamma,\Omega)$ is highly
localized by a stiff potential with respect to $\beta$.  For
specificity, consider
$V(\beta,\gamma)=u(\beta)+v(\gamma)$.  Then $H\approx H_\beta+\beta_0^{-2}
H_{\gamma\Omega}$, where $H_\beta=-\Deltahat+u(\beta)$ and
$H_{\gamma\Omega}=\Lambdahat^2+\beta_0^2 v(\gamma)$.  The separated eigenfunctions
$\Psi(\beta,\gamma,\Omega)= f(\beta)\psi(\gamma,\Omega)$ satisfy $H_\beta f(\beta) =
\varepsilon_\beta f(\beta)$ and 
$H_{\gamma\Omega}\psi(\gamma,\Omega)=\varepsilon_{\gamma\Omega}\psi(\gamma,\Omega)$.
Note that the angular
problem is thus of the
form~(\ref{eqn-gammarotor}), with $V(\gamma)\equiv
\beta_0^2v(\gamma)$.  The total energy eigenvalues $E$, defined by $H\Psi(\beta,\gamma,\Omega)=E\Psi(\beta,\gamma,\Omega)$,
are obtained additively as
$E=\varepsilon_\beta+\beta_0^{-2}\varepsilon_{\gamma\Omega}$.
Therefore, for fixed $\beta$ excitation (\textit{e.g.}, the ground
state for the $\beta$ problem), the eigenvalues of the angular problem
directly give the energy spectrum.  These arguments apply only in the
limit of stiff $\beta$ confinement, and finite $\beta$ softness may be
expected to lead to $\beta$-$\gamma$
coupling~\cite{caprio2005:axialsep}.

(2)~Alternatively, for $V(\beta,\gamma)=u(\beta)+v(\gamma)/\beta^2$,
the Bohr Hamiltonian eigenproblem is exactly
separarable~\cite{jean1960:transition-model}.  In this case, $H=
H_\beta+\beta^{-2} H_{\gamma\Omega}$, where 
$H_\beta=-\Deltahat+u(\beta)$ and now $H_{\gamma\Omega}=\Lambdahat^2+
v(\gamma)$.  The separated eigenfunctions $\Psi(\beta,\gamma,\Omega)=
f(\beta)\psi(\gamma,\Omega)$ satisfy $(H_\beta
+\beta^{-2}\varepsilon_{\gamma\Omega})f(\beta) = E f(\beta)$ and
$H_{\gamma\Omega}\psi(\gamma,\Omega)=\varepsilon_{\gamma\Omega}\psi(\gamma,\Omega)$.
Note that the angular problem is of the form~(\ref{eqn-gammarotor}) with
$V(\gamma)\equiv v(\gamma)$.  The eigenvalue
$\varepsilon_{\gamma\Omega}$ from the angular problem now appears in
the $\beta$ equation as a ``centrifugal'' coefficient, \textit{i.e.},
multiplying $\beta^{-2}$.  It therefore
enters indirectly into the total eigenvalue $E$, through the $\beta$
eigenproblem, rather than directly giving the energy spectrum.

\section{Mikhailov relations}
\label{app-mikhailov}

This appendix adapts the leading-order
$\Delta K=2$ bandmixing relations~(\ref{eqn-mikhailov-DeltaK2})
and~(\ref{eqn-mikhailov-DeltaK2-IME}) to the form required for the
analysis of Figs.~\ref{fig-gammarotor-axial-mikhailov} and~\ref{fig-gammarotor-triax-mikhailov}.
For $K$-\textit{decreasing} transitions (\textit{e.g.},
$\gamma\rightarrow g$ and $\gamma\gamma_4\rightarrow\gamma$), in terms
of $B(E2)$ reduced transition probabilities,
\begin{multline}
\label{eqn-mikhailov-decr}
B(E2;K_iJ_i\rightarrow K_fJ_f) = \sigma_i^2 
\tcg{J_i}{K_i}{2}{-2}{J_f}{K_f}^2
\ifproofpre{\\\times}{}
M_1^2\bigl[1+a[J_f(J_f+1)-J_i(J_i+1)]\bigr]^2,
\end{multline}
with normalized \textit{positive} slope parameter $a=-M_2/M_1$.  Thus, the intrinsic
matrix element is extracted as
\begin{equation}
\label{eqn-mikhailov-decr-IME}
\tme{K_f}{\calM'}{K_i}=M_1[1-4(K_f+1)a].
\end{equation}
Similarly, for $K$-\textit{increasing} transitions (\textit{e.g.},
$\gamma\gamma_0\rightarrow\gamma$),
\begin{multline}
\label{eqn-mikhailov-incr}
B(E2;K_iJ_i\rightarrow K_fJ_f) = \sigma_i^2 
\tcg{J_i}{K_i}{2}{+2}{J_f}{K_f}^2
\ifproofpre{\\\times}{}
M_1^2\bigl[1+a[J_f(J_f+1)-J_i(J_i+1)]\bigr]^2,
\end{multline}
where now the \textit{positive} slope parameter is $a=+M_2/M_1$, and
thus the intrinsic matrix element is extracted as
\begin{equation}
\label{eqn-mikhailov-incr-IME}
\tme{K_f}{\calM'}{K_i}=M_1[1+4(K_i+1)a].
\end{equation}

\section{Wave function probability relations}
\label{app-prob}

In this appendix, expressions are given for the probability
distribution $P(\gamma)$, with respect to the $\gamma$ coordinate, and
the decomposition $P_K$, with respect to the $K$ quantum number, for a
wave function $\psi(\gamma,\Omega)$.  Note that the volume element for
the coordinates $(\gamma,\Omega)$ is given by $\abs{\sin
  3\gamma}\,d\gamma\,d\Omega$.

The probability distribution $P(\gamma)$  is obtained by 
integration over Euler angles, as
\begin{equation}
\label{eqn-Pgamma-defn}
P(\gamma)=\abs{\sin 3\gamma} \int \abs{\psi(\gamma,\Omega)}^2 \, d\Omega,
\end{equation}
and thus, in terms of the (real) coefficient functions $F_K(\gamma)$
appearing in (\ref{eqn-expansion-F}), 
\begin{equation}
\label{eqn-Pgamma-F}
P(\gamma)=\frac{16\pi^2}{2L+1}\abs{\sin3\gamma}\sum_{\substack{K=0\\\text{even}}}^L
[F_{K}(\gamma)]^2.
\end{equation}
The angular integration has been carried out using the orthogonality integral for the 
$\scrD$
functions~\cite{edmonds1960:am}, which gives
$\int \xi^{(L')\,*}_{K'M'}  (\Omega)\xi^{(L)}_{KM} (\Omega) \, d\Omega = [(16 \pi^2)/(2L+1)]
\linebreak[0]\delta_{L'L} 
\linebreak[0]\delta_{K'K} 
\linebreak[0]\delta_{M'M}$, unless $K=0$ with $L$
odd, in which case the integral vanishes~\cite{caprio2009:gammaharmonic}.  For the eigenfunctions
$\psi_{LiM}(\gamma,\Omega)$ obtained with respect to the $\grpsochain$
basis, the known quantities are the diagonalization
coefficients $a_{Lij}$ appearing in~(\ref{eqn-diag-expansion}) and the
functions $F_{LiK}(\gamma)$ in the representation of the $\grpsochain$ spherical harmonics
\begin{equation}
\label{eqn-expansion-Psi-F}
\Psi_{LiM}(\gamma,\Omega)=\sum_{\substack{K=0\\\text{even}}}^L
F_{LiK}(\gamma) \xi^{(L)}_{KM}(\Omega),
\end{equation}
where we again use a counting index to label the spherical harmonics.
\setlength{\footnotesep}{30pt}
In terms of these,\fnPgamma
\begin{multline}
\label{eqn-Pgamma-sum}
P_{Li}(\gamma)=\frac{16 \pi^2}{2L+1}\abs{\sin3\gamma}
\ifproofpre{\\\times}{}
\sum_{\substack{K=0\\\text{even}}}^L\sum_{jk}a_{Lij}a_{Lik}
F_{LjK}(\gamma)F_{LkK}(\gamma).
\end{multline}

The contribution of each $K$ value 
to $\psi(\gamma,\Omega)$, integrated over $\gamma$, is 
\begin{equation}
\label{eqn-PK-defn}
P_K=
\frac{16\pi^2}{2L+1}\int_0^{\pi/3}
[F_{K}(\gamma)]^2
\,\sin3\gamma \,d\gamma.
\end{equation}
For the functions
$\psi_{LiM}(\gamma,\Omega)$, represented by $a_{Lij}$ coefficients
with respect to the $\grpsochain$
basis, these probabilities may be computed as
\begin{multline}
\label{eqn-PK-sum}
P_{Li;K}
=
\frac{16\pi^2}{2L+1}
\sum_{jk}a_{Lij}a_{Lik}
\ifproofpre{\\\times}{}
\int_0^{\pi/3} F_{LjK}(\gamma)F_{LkK}(\gamma) \,\sin 3\gamma \,
d\gamma.
\end{multline}

\vfil

%***************************************************************************
% bibliography
%***************************************************************************

%%\clearpage

% Bibliography created with apsrevm.bst
\providecommand{\APSLONG}{}
\providecommand{\ELSEVIER}{}
\newcommand{\identity}[1]{{#1}}

%bibliography{master,mc,theory,expt,books,proc,misc,gammatriax2}

\end{document}